\documentclass[lettersize,journal]{IEEEtran}
        
\usepackage{color,amsmath,amssymb,amsthm,epsfig,mathrsfs,cite,bm,graphics}
\usepackage{algpseudocode}
\usepackage{algorithmicx}
\usepackage{algorithm} 
\usepackage[utf8]{inputenc}
\usepackage{cite}
\usepackage{amsmath}
\DeclareMathOperator*{\argmax}{argmax}

\usepackage{amsfonts}
\usepackage{amssymb}
\usepackage{placeins}
\usepackage{graphicx}
\usepackage{bbm,bm}
\usepackage{graphicx,graphics,subcaption}
\usepackage{tikz}
\usepackage{pgf}
\usepackage{pgfplots}
\usepackage{hyperref}
\usepackage[all]{hypcap} 
\usepackage{float}
\usepackage{tabularx}

\usetikzlibrary{automata, positioning, arrows,shapes.multipart,fit}
        
\usepackage{todonotes}
\presetkeys{todonotes}{color=blue!20}{}
\usepackage{pgfplots}
\usepackage{pgfplotstable}

\usepackage{amssymb}
\usepackage{pifont}
\newcommand{\xmark}{\ding{55}}%

\title{A DRL Approach for RIS-Assisted Full-Duplex UL and DL Transmission: Beamforming, Phase Shift and Power Optimization}
\author{Nancy Nayak, Sheetal Kalyani,~\IEEEmembership{ Member,~IEEE,} and Himal A. Suraweera,~\IEEEmembership{Senior Member,~IEEE}

\thanks{This research work was partially supported by the Ministry of Electronics 
 \& Information Technology (Government Of India), under Project SP21221155EEMEIT008073.}
 
\thanks{N. Nayak and S. Kalyani are with the Department of Electrical Engineering, Indian Institute of Technology Madras, Chennai, India (e-mail: ee17d408@smail.iitm.ac.in, skalyani@ee.iitm.ac.in).}

\thanks{H. A. Suraweera is with the Department of Electrical and Electronic Engineering, University of Peradeniya, Peraeniya 20400, Sri Lanka (e-mail: himal@eng.pdn.ac.lk).}

}

\begin{document}

    \maketitle

\begin{abstract}
We propose a deep reinforcement learning (DRL) approach for a full-duplex (FD) transmission that predicts the phase shifts of the reconfigurable intelligent surface (RIS), base station (BS) active beamformers, and the transmit powers to maximize the weighted sum rate of uplink and downlink users. Existing methods require channel state information (CSI) and residual self-interference (SI) knowledge to calculate exact active beamformers or the DRL rewards, which typically fail without CSI or residual SI. Especially for time-varying channels, estimating and signaling CSI to the DRL agent is required at each time step and is costly. We propose a two-stage DRL framework with minimal signaling overhead to address this. The first stage uses the least squares method to initiate learning by partially canceling the residual SI. The second stage uses DRL to achieve performance comparable to existing CSI-based methods without requiring the CSI or the exact residual SI. Further, the proposed DRL framework for quantized RIS phase shifts reduces the signaling from BS to the RISs using $32$ times fewer bits than the continuous version. The quantized methods reduce action space, resulting in faster convergence and $7.1\%$ and $22.28\%$ better UL and DL rates, respectively than the continuous method.
\end{abstract}

\section{Introduction}
The research on fifth-generation (5G) wireless systems has focused on achieving excellent throughput, ultra-low latency, high reliability, global coverage, and connectivity. Reconfigurable intelligent surface (RIS) \cite{di2020smart} has drawn attention due to its impressive massive MIMO-like gains with low cost and less energy consumption\cite{abeywickrama2020intelligent} and its adaptability in technologies such as Millimeter-Wave multiple input multiple output (MIMO) communication, index modulation, non-orthogonal multiple access, UAV communication, physical layer security, cognitive radio \cite{wang2020jointpelian,gopi2020intelligent,cheng2021downlink,li2020reconfigurable} and so on. Different studies have been performed to understand the effectiveness of RIS, such as the analysis of outage probability, SINR, capacity, secrecy performance, energy efficiency, etc. \cite{charishma2021outage,kammoun2020asymptotic,jayalal2022sinr,du2021capacity,huang2019reconfigurable}. The initial works on RIS focused on channel estimation and passive beamforming (RIS phase shift) in a half-duplex (HD) setup using different methods like alternating optimization, semi-definite relaxation (SDR) \cite{wu2019intelligent}, fix point iteration \cite{yu2019miso}, minimum variance unbiased estimator \cite{jensen2020optimal}, etc. Since full-duplex (FD) communication attains better spectral efficiency than HD system \cite{sabharwal2014}, the performance of RIS is studied in an FD setup. However, it is assumed that the strong line-of-sight component of the self-interference (SI) channel is estimated and removed during the cancellation process employed at the base station (BS) to exploit the power of the FD setup. However, when the SI cancellation technique is inadequate or inefficient, the power of the residual SI is high. Many works on RIS in an FD setup either do not consider SI \cite{pan2021full} or assume that the SI is effectively suppressed up to the background noise ﬂoor \cite{bharadia2013full} by several SI cancellation techniques such as passive suppression, analog and digital cancellations, etc. \cite{riihonen2011,everett2014passive,smida2023full}.   

In a multi-RIS FD setup, \cite{cai2021intelligent} minimizes the weighted sum of the transmit powers of the BS and uplink user equipment (ULue) by jointly optimizing the phase shifts at the RISs, subject to the rate constraints of ULue and downlink user equipment (DLue). A study investigated a point-to-point communication system supported by a RIS with only one reflected link where two devices were operating in FD mode \cite{yang2021optimal}. In \cite{zhang2020sum}, the authors explored optimizing continuous and discrete RIS phase shifts in a communication system involving direct and reflected links. The authors of \cite{nguyen2021cooperative} examined a bidirectional FD communication system that employed multiple RISs and analyzed outage probability and ergodic capacity; they also demonstrated that the adverse effects of SI could be mitigated by increasing the size of the RISs. The authors of \cite{elhattab2021reconfigurable} show that the FD C-NOMA has more resistance to the residual SI effect with the assistance of RIS than without RIS. All the works \cite{cai2021intelligent,yang2021optimal,zhang2020sum,nguyen2021cooperative,elhattab2021reconfigurable,perera2022sum,peng2021multiuser} assume that the SI is eliminated effectively, and therefore, the residual SI is very low. However, the residual SI need not be the case in practical scenarios. The authors in \cite{zhang2015full} clearly state that achieving low residual SI may involve complex methods and not always be practically possible.

\ifCLASSOPTIONtwocolumn
\begin{table*}[t]
\caption{Key literature studying RIS-assisted FD communication systems. Q: quantization, G: grouping.}
\label{tab:allmethods}
  \begin{center}
    \begin{tabular}{|p{1.5cm}|p{0.3cm}|p{1.0cm}|p{2.0cm}| p{1.2cm} |p{1.0cm}|p{1.0cm} | p{5.4cm}|p{0.4cm}|}
    \hline
    \textbf{Paper} & \textbf{FD} &\textbf{BS-UE link} & \textbf{No knowledge of CSI and residual SI} & \textbf{Transmit powers} & \textbf{Beam-formers} & \textbf{RIS phase shift} & \textbf{Technique}& \textbf{Q,G}\\
    \hline
    \cite{cai2021intelligent}, 2021  & \checkmark & \checkmark  & \xmark & \checkmark & \xmark & \checkmark & Block Coordinate Descent (BCD) & \xmark\\
    \hline
    \cite{peng2021multiuser}, 2021 & \checkmark & \xmark  & \xmark & \xmark & \checkmark & \checkmark & (i) BCD, (ii) Minorization-Maximization (MM) & \xmark \\
    \hline
    \cite{taha2020deep}, 2021 & \xmark& \xmark & - & \xmark & \xmark & \checkmark & Deep $Q$-learning & \xmark\\
    \hline
    \cite{feng2020deep}, 2021 & \xmark & \xmark & - & \xmark & \checkmark & \checkmark  & RIS phase shift by DDPG, beamforming vector by MRT & \xmark\\
    \hline
    \cite{lin2020deep}, 2020 & \xmark & \checkmark & - &  \xmark & \checkmark & \checkmark & RIS phase shift by DDPG, beamforming by convex optimization & \xmark\\
    \hline
    \cite{faisal2021deep}, 2022 &  \checkmark & \checkmark & \xmark & \xmark & \checkmark & \checkmark & RIS phase shift by DRL, beamformer by MRC/MRT/ZF & \xmark\\
    \hline
    \cite{subhash2022max}, 2022 & \xmark & \xmark & \xmark &  \checkmark & \checkmark & \checkmark & Beamforming by MRC, matrix lifting method for phase shift and geometric programming for power allocation. Heuristic algorithm for quantized RIS phase shift. & Q \\
    \hline
    \cite{shekhar2022instantaneous}, 2022 & \xmark & \checkmark & \xmark & \xmark & \checkmark & \checkmark & Quantized phase shifts by MPSO and PSO, beamforming by MRC. & Q \\
    \hline
    This work & \checkmark & \checkmark & \checkmark & \checkmark & \checkmark & \checkmark & Complete DRL solution with DDPG algorithm & Q,G\\
    \hline
\end{tabular}
\end{center}
\end{table*}
\fi

The methods that assume low residual SI use active beamformers to improve the UL and/or DL rates \cite{alwazani2020intelligent, abeywickrama2020intelligent, luo2021spatial}. On the other hand, the methods that assume a significantly high residual SI use the beamformers to cancel the residual SI at the BS \cite{faisal2021deep, mohammadi2015full}. However, to design these beamformers, one requires accurate knowledge of the channel state information (CSI) and the residual SI. The assumption of perfect knowledge of the CSI may not be practical. To the best of our knowledge, all the existing works need CSI to design their beamformers. However, there have been attempts to use statistical CSI instead of instantaneous CSI in the context of performance and outage analysis, robust transmission in vehicular communication, etc., to reduce the amount of signaling overhead for updating the RIS phase shifts \cite{zhi2021statistical,chen2021robust}. The authors of \cite{subhash2022optimal} design the continuous RIS phases by minimizing the signal-to-interference-plus-noise-ratio (SINR) that depends on the statistical CSI. Using only statistical CSI, \cite{shekhar2022instantaneous} solves for quantized RIS phases using particle swarm optimization to minimize the outage probability and maximize the ergodic rate. In our work, without using any knowledge of CSI (statistical or instantaneous), we propose a solution that maximizes the weighted sum of the UL and DL rates. Also, as we do not assume any knowledge of SI, the solution is expected to handle high residual SI power.

The works mentioned above \cite{
wu2019intelligent,wu2019beamforming,yu2019miso,jensen2020optimal,pan2021full,cai2021intelligent,perera2022sum,peng2021multiuser,alwazani2020intelligent, abeywickrama2020intelligent, luo2021spatial,subhash2022optimal,yang2021optimal,zhang2020sum,elhattab2021reconfigurable,nguyen2021cooperative} solve the non-convex optimization problem of finding optimal beamformer and/or phase shifts using either the traditional mathematical alternating optimization method or the block coordinate descent method. These solutions are computationally intensive. With more UL and DL users, RIS elements, and unknown channel models, the prediction of reflection matrices of the RIS becomes more complicated. Unknown communication scenarios, such as hardware impairments, add to the complexity of predicting reflection matrices, beamforming vectors, and transmit power. In our opinion, such complex high-dimensional problems can be easily addressed by exploiting the function approximation capabilities of deep networks, such as deep supervised learning. However, this required a large set of collected and labeled data, which can be challenging to acquire. 

In contrast, deep reinforcement learning (DRL) does not need labeled data. DRL combines the advantages of function approximation by deep learning and policy-making by reinforcement learning. In \cite{taha2020deep}, authors predict RIS phases and do not consider beamforming, as the BS has a single antenna transmitter. In \cite{feng2020deep}, the authors use a DRL framework to predict the RIS phases in a DL setup, where the beamforming vectors are designed using maximum ratio transmission (MRT) to improve the DL rates. A convex optimization-driven deep deterministic policy gradient (DDPG) algorithm is proposed in \cite{lin2020deep} to solve a power minimization problem in a RIS-assisted multiple-input-single-output (MISO) system, subject to the receiver’s worst-case data rate requirement, and the RIS’s worst-case power budget constraint. The authors of \cite{faisal2021deep} consider FD mode in a RIS-assisted multi-antenna BS system and alternatively optimize phase shifts and beamformers. Though the RIS phase shifts are calculated using the DRL method, the beamformers are calculated using the conventional optimization method, which still needs complete CSI information. In a DRL setup, when the residual SI is high, the signal received at BS will have a high interference level, and hence, the algorithm will not learn effectively. If the algorithm cannot learn well, the design of the beamformers will be inaccurate, and one cannot cancel the residual SI. To solve this coupled problem of beamformer design and residual SI cancellation, it is essential to iteratively cancel the residual SI and solve for the beamformers at that instant. We propose a least square-based SI cancellation method; in cooperation with the DRL method, this can handle high residual SI power without costly SI mitigation schemes. This online solution maximizes the weighted sum of the UL and DL rates. During optimization, it allows us to altogether avoid the CSI and the knowledge of residual SI, reducing the signaling overhead to the BS.

Recently, researchers have focused on quantized phase shifts at RIS so that the signaling required between BS and RISs is less \cite{shekhar2022instantaneous, wu2019beamforming}. The work in \cite{subhash2022max} solves for the continuous RIS phase shifts, beamforming, and power allocation by alternating optimization to maximize the minimum SINR, whereas to solve the quantized RIS phase shift values, the authors use a heuristic algorithm. Therefore, the traditional solution methods used for continuous phase shifts must be changed to solve quantized phases. However, we show that by slightly modifying our architecture, the proposed DRL algorithm can predict quantized RIS phase shifts and performs better than the algorithm with continuous RIS phase shifts. To reduce the signaling further, we group the passive elements of the RIS and only use a single phase shift to describe the group.

In summary, we introduce a novel architecture consisting of a feature extractor and multiple sub-networks, whose outputs are transformed appropriately to represent the RIS phase shifts, beamformers, and transmit powers. Instead of picking the exact RIS phase shifts, beamformers, and transmit powers, noise is added to RIS phase shifts and beamformers for better exploration and faster convergence. Table \ref{tab:allmethods} highlights the distinctions between our method and prior art. Our key contributions are as follows.
\begin{itemize}
    \item We propose a two-stage learning algorithm, which assumes the absence of a) a good SI mitigation scheme and b) the costly CSI overhead but performs almost as well as perfect CSI-based semi-oracle DRL methods.
    \item To overcome the challenge of SI, a least square-based method is proposed when a good estimate of SI is not present. 
    \item The proposed method needs minimum signaling overhead as it does not need any CSI to be communicated to the learning algorithm. 
    \item To further reduce the signaling, we propose a DRL framework that can handle pseudo-discrete action space to predict a) quantized RIS phases and b) reduced phase information with one phase information per group. The quantized phase DRL method has $32$ time lesser signaling than the continuous phase DRL method, with better convergence. 
\end{itemize}

    \section{System Model and Problem Formulation}
Consider an FD system consisting of a BS, one ULue, and one DLue. The BS has a uniform linear antenna (ULA) array on the y-axis with $M_t^{}$ transmit antenna elements and $M_r^{}$ receive antenna elements and operates in the FD mode. Both ULue and DLue are single-antenna HD user nodes. The direct links between the BS and the user equipment (ULue and DLue) are assumed to be weak due to blockages. Therefore, two RISs are placed at the two ends of the area to boost the communication between BS and the users (ULue and DLue), as shown in Fig. \ref{fig:metrocity}. The ULue and DLue are assumed not to have a line-of-sight (LoS). Both the RISs are deployed as Uniform Planar Antenna (UPA) in the xz plane; RIS1 and RIS2 have $N_1^{}=N_{1h}^{}N_{1v}^{}$ and $N_2^{}=N_{2h}^{}N_{2v}^{}$ reflecting elements, respectively. The diagonal phase shift matrices $\mathbf{\Theta}_U^{}$ and $\mathbf{\Theta}_D^{}$ of the two RISs are given as:
\ifCLASSOPTIONtwocolumn
\begin{equation}
\begin{aligned}
    \mathbf{\Theta}_{U}^{}&=\text{diag}\{\Bar{\Theta}_U^{}\} = \text{diag}\{\phi_{U1}^{}\, \dots\, \phi_{UN_1}^{}\}, \text{ and } \\
    \mathbf{\Theta}_{D}^{}&=\text{diag}\{\Bar{\Theta}_D^{}\}=\text{diag}\{\phi_{D1}^{}\, \dots\, \phi_{DN_2}^{}\}
\end{aligned}
\end{equation}
\fi
with $\phi_n=e^{j\theta_n^{}}$. Let $s_U^{}$ denote the transmit signal from the ULue, and $p_U^{}>0$ denote the transmit power of the ULue. Let $s_D^{}$ denote the transmit signal of the BS, and $p_A^{}>0$ denotes the transmit power of the BS. Let $\mathbf{n}_A^{}\in \mathbb{C}^{M_r^{}\times 1}$ be the i.i.d. complex Gaussian noise with zero mean and variance $\sigma_A^2$. Let $\mathbf{f}_{IU}^{}\in \mathbb{C}^{N_1\times 1}$, $\mathbf{F}_{AI}^{}\in \mathbb{C}^{M_r^{}\times N_1^{} }$, $\mathbf{h}_{AU}^{}\in \mathbf{C}^{M_r^{}\times 1}$, $\mathbf{G}_{IA}^{}\in\mathbb{C}^{N_2^{}\times M_t^{}}$, and $\mathbf{g}_{IU}^{}\in\mathbb{C}^{N_2^{}\times 1}$ be the channels from ULue to RIS1, from RIS1 to the BS, from ULue to the BS, from the BS to RIS2, and from ULue to RIS2, respectively. The phase shift matrix at RIS1 is given by $\mathbf{\Theta}_U^{}\in\mathbb{C}^{N_1^{}\times N_1^{}}$ and $\mathbf{H}_{AA}^{}\in\mathbb{C}^{M_r^{} \times M_t^{}}$ is the channel between the transmit and the receive antennas at the BS. The path loss of $\mathbf{G}_{IA}^{}$ and $\mathbf{F}_{AI}^{}$ is much higher than $\mathbf{H}_{AA}^{}$, so the reflecting SI from the RISs to the BS are ignored \cite{xu2020resource}. Therefore, the received signal at the BS is given by
\ifCLASSOPTIONtwocolumn
\begin{equation}
  \begin{aligned}
    \label{eq:UL}
    y_A^{} &= \mathbf{w}_R^{} \mathbf{h}_{AU}^{}\sqrt{p_U^{}}s_U^{}+\mathbf{w}_R^{}\mathbf{F}_{AI}^{}\mathbf{\Theta}_U^{}\mathbf{f}_{IU}^{}\sqrt{p_U^{}}s_U^{} \\ &+\mathbf{w}_R^{}\mathbf{G}_{IA}^{T}\mathbf{\Theta}_D^{}\mathbf{g}_{IU}^{}\sqrt{p_U^{}}s_U^{} +\mathbf{w}_R^{}\mathbf{H}_{AA}^{}\mathbf{w}_T^{}\sqrt{p_A^{}}s_D^{} \\ &+ \mathbf{w}_R^{}n_A^{},
\end{aligned}  
\end{equation}
\fi
\ifCLASSOPTIONtwocolumn
\begin{figure}[!t]
    \centering
    \input{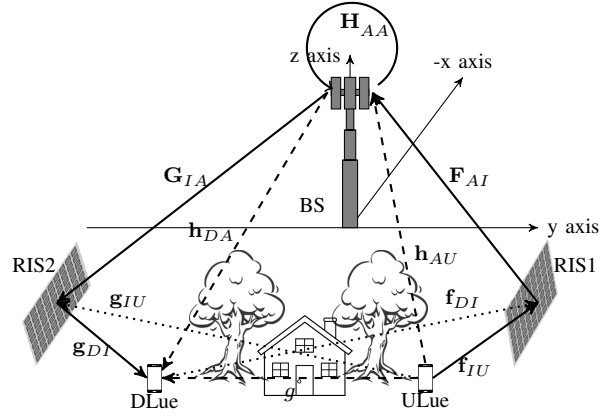}
    \caption{An FD communication scenario setup with one ULue, one DLue, and two RISs to facilitate communication when the users are not in LoS with the BS.}
    \vspace{-2mm}
    \label{fig:metrocity}
\end{figure}
\fi
where $\mathbf{w}_T^{}\in\mathbb{C}^{M_t^{}\times 1}$ and $\mathbf{w}_R^{}\in\mathbb{C}^{1\times M_r^{}}$ are the transmit and the receive beamforming vector at the BS, respectively, assuming BS has full digital transmit and receive beamforming. Let $n_D^{}\in \mathbb{C}$ be the i.i.d. complex Gaussian noise with zero mean and variance $\sigma_D^2$. Let, $\mathbf{g}_{DI}^{}\in\mathbb{C}^{1\times N_2^{}}$, $\mathbf{h}_{DA}^{}\in \mathbf{C}^{1\times M_t^{}}$, $\mathbf{f}_{DI}\in\mathbb{C}^{1\times N_1}$ and $g\in\mathbb{C}^{1\times 1}$ are the channels from RIS2 to DLue, from the BS to DLue, and from ULue to DLue, respectively. The phase shift matrix at RIS2 is given by $\mathbf{\Theta}_D^{}\in\mathbb{C}^{N_2^{}\times N_2^{}}$. Then, the signal received by the DLue is
\ifCLASSOPTIONtwocolumn
\begin{equation}
\begin{aligned}
\label{eq:DL}
    y_D^{} &=\mathbf{h}_{DA}^{}\mathbf{w}_T^{}\sqrt{p_A^{}}s_D^{} +\mathbf{g}_{DI}^{}\mathbf{\Theta}_D^{}\mathbf{G}_{IA}^{}\mathbf{w}_T^{}\sqrt{p_A^{}}s_D^{} 
        \\
        &+\mathbf{f}_{DI}^{}\mathbf{\Theta}_U^{}\mathbf{F}_{AI}^{T}\mathbf{w}_T^{}\sqrt{p_A^{}}s_D^{} +\mathbf{g}_{DI}^{}\mathbf{\Theta}_D^{}\mathbf{g}_{IU}^{}\sqrt{p_U^{}}s_U^{} \\ &+\mathbf{f}_{DI}^{}\mathbf{\Theta}_U^{}\mathbf{f}_{IU}^{}\sqrt{p_U^{}}s_U^{}
        + g\sqrt{p_U^{}}s_U^{}+n_D^{}.
\end{aligned}
\end{equation}
\fi

The SINRs at the BS, $\gamma_{BS}$ and at the DL user, $\gamma_{DL}$ are given by 
\ifCLASSOPTIONtwocolumn
\begin{equation}
    \begin{aligned}
    \gamma_{BS}^{}  &= \frac{p_U^{}||\mathbf{w}_R^{} (\mathbf{h}_{AU}^{}+\mathbf{F}_{AI}^{}\mathbf{\Theta}_U^{}\mathbf{f}_{IU}^{}+\mathbf{G}_{IA}^{T}\mathbf{\Theta}_D^{}\mathbf{g}_{IU}^{})||^2}{p_A^{}||\mathbf{w}_R^{}\mathbf{H}_{AA}^{}\mathbf{w}_T^{}||^2+\mathbf{w}_R^2\sigma_A^2}\text{, and }\\
    \gamma_{DL}^{} &= \frac{p_A^{}||  (\mathbf{h}_{DA}^{}\mathbf{w}_T^{}+\mathbf{g}_{DI}^{}\mathbf{\Theta}_D^{}\mathbf{G}_{IA}^{}\mathbf{w}_T^{}+\mathbf{f}_{DI}^{}\mathbf{\Theta}_U^{}\mathbf{F}_{AI}^{T}\mathbf{w}_T^{})||^2}{p_U^{}||(g+\mathbf{g}_{DI}^{}\mathbf{\Theta}_D^{}\mathbf{g}_{IU}^{}+\mathbf{f}_{DI}^{}\mathbf{\Theta}_U^{}\mathbf{f}_{IU}^{})||^2+\sigma_D^2}.
\end{aligned}
\end{equation}
\fi
Accordingly, the data rates at the DLue and the BS are given by
\ifCLASSOPTIONtwocolumn
\begin{equation}
    \begin{aligned}
    \label{}
    r_{DL} &= \log_2 \left( 1 + \gamma_{DL}^{} \right), \text{ and }
    r_{BS} &= \log_2 \left( 1 + \gamma_{BS}^{} \right).
    \label{eqn:datarate}
\end{aligned}
\end{equation}
\fi

In this work, the weighted sum of UL and DL data rate is maximized by optimizing jointly over the RIS phase shifts $\mathbf{\Theta}_{D}$, $\mathbf{\Theta}_{U}$, the transmit and receive beamformers $\mathbf{w}_T$ and $\mathbf{w}_R$ and the transmit power of BS and ULue given by $p_A$ and $p_U$, respectively. Therefore, the optimization problem is formulated as
\begin{equation}
    \begin{aligned}
    \label{eq:optobjective}
    \mathcal{P}_1^{}: \quad  \underset{\mathbf{\Theta}_{D}^{}, \mathbf{\Theta}_{U}^{}, \mathbf{w}_T^{}, \mathbf{w}_R^{}, p_A^{}, p_U^{}}{\max}
    & {\delta r_{BS}^{}+(1-\delta)r_{DL}^{}} \\
    \text{s.t. }  0 \leq p_A^{} \leq & p_{A}^{max}, \quad 0 \leq p_U^{} \leq p_{U}^{max}, \\ |\phi_n^{}|=1, 1\leq n &\leq N_1^{}, 1\leq n \leq N_2^{},
\end{aligned}
\end{equation}
where $\delta$ is a hyperparameter that regulates the weights of UL and DL data rates in the optimization objective; the maximum transmit power of ULue and BS are given by $ p_U^{max}$ and $ p_A^{max}$, respectively. In the context of RIS, the optimization problem presents significant challenges and cannot be effectively addressed in its current formulation. The optimization problems are typically relaxed, which deviates from the practical scenario. Traditional optimization methods usually break the big problem down into smaller sub-problems, such as channel estimation, beamformer calculation, and the RIS phase shift estimation. The methods are computationally expensive, especially when the number of variables is large. Furthermore, existing works either assume an excellent SI mitigation method, resulting in negligible residual SI power. Alternatively, if the residual SI is high, they assume knowledge of the residual SI and the CSI to design the beamformers to cancel the residual SI. In this work, we propose a \textit{two-stage learning method} in which the first stage of each time step involves an SI-cancellation technique that reduces a part of SI to start the learning process. The second stage exploits the potential of DRL to find the beamformers, RIS phases, and transmit powers. However, one needs to smartly formulate the actions, observations, and rewards for the DRL algorithm to make the agent learn.

\section{Proposed two-stage learning method }
\subsection{First stage: residual SI-cancellation Method}
\label{sec:SI}
The main challenge in the FD communication system is the SI imposed by the transmit antenna on the receive antenna of the BS. It is difficult for the DRL agent to learn anything if the received signal has too much interference due to the high residual SI. Therefore, canceling the SI is vital in enabling the agent to learn. The signal due to SI can be canceled with an estimate of $\mathbf{w}_R^{}\mathbf{H}_{AA}^{}\mathbf{w}_T^{}$ denoted by $\hat{h}$ as 
\ifCLASSOPTIONtwocolumn
\begin{equation}
    \begin{aligned}
    \label{eq:SIcancellation}
    \Tilde{y}_A^{} &= \mathbf{w}_R^{} \mathbf{h}_{AU}^{}\sqrt{p_U^{}}s_U^{}+\mathbf{w}_R^{}\mathbf{F}_{AI}^{}\mathbf{\Theta}_U^{}\mathbf{f}_{IU}^{}\sqrt{p_U^{}}s_U^{} \\ &+ \mathbf{w}_R^{}\mathbf{G}_{IA}^{T}\mathbf{\Theta}_D^{}\mathbf{g}_{IU}^{}\sqrt{p_U^{}}s_U^{} + (\mathbf{w}_R\mathbf{H}_{AA}^{}\mathbf{w}_T^{}-\hat{h})\sqrt{p_A^{}}s_D^{} \\& + \mathbf{w}_R^{}n_A^{}.
\end{aligned}
\end{equation}
\fi
We show two ways to find an estimate $\hat{h}$ for canceling the SI.

\subsubsection{Least square-based SI-cancellation} First, we propose a simple least square-based SI-cancellation (LSSIC) method using a single pilot signal. The scalar $\hat{h}$ is estimated at every epoch by sending a pilot signal from the transmitter to the receiver antenna at the BS. Let a pilot signal $s_D^p \in \mathcal{C}$ be transmitted from the transmit antenna at BS. The corresponding received signal $y_A^p\in\mathcal{C}$ at the receiver antenna of BS can be expressed as
\begin{align}
\label{eq:LSSIC}
    y_A^p &= \mathbf{w}_R^{}\mathbf{H}_{AA}^{}\mathbf{w}_T^{}\sqrt{p_A^{}}s_D^p + v_A^{},
\end{align}
where $v_A^{}\in\mathcal{C}$ is the AWGN and the scalar $\mathbf{w}_R^{}\mathbf{H}_{AA}^{}\mathbf{w}_T^{}$ needs to be estimated. Note that $\mathbf{w}_T^{}\in\mathbb{C}^{M_t^{}\times 1}$ and $\mathbf{w}_R^{}\in\mathbb{C}^{1\times M_r^{}}$. To estimate $\mathbf{w}_R\mathbf{H}_{AA}^{}\mathbf{w}_T^{}$, we need to minimize the error $J(h)$ where
\begin{equation}
    \begin{aligned}
    J(h) &= \overline{(y_A^p - \sqrt{p_A^{}}\,s_D^p\,h)}(y_A^p - \sqrt{p_A}\,s_D^p\,h), \\
    &=\overline{y_A^p}\,y_A^p - 2\sqrt{p_A^{}}\,h\,\overline{y_A^p}\,s_D^p + p_A^{}\,h^2\,\overline{s_D^p}\,s_D^p.
\end{aligned}
\end{equation}
where $\Bar{x}$ denotes the conjugate transpose of $x$. The optimal $h$ is found by taking the derivative of $J(h)$ with respect to $h$ and equating it to zero to obtain $\hat{h}$,
\begin{equation}
    \begin{aligned}
    \label{eq:HSIC}
    \frac{\partial J(h)}{\partial h} &= 0 - 2\sqrt{p_A^{}}\,\overline{y_A^p}\,s_D^p +2p_A^{}\,h\,\overline{s_D^p}\,s_D^p = 0, \\ \text{so, }
    \hat{h} &= \frac{1}{\sqrt{p_A}}(\overline{s_D^p}s_D^p)^{-1}\,\overline{s_D^p}\,y_A^p.
\end{aligned}
\end{equation}
The derived $\hat{h}$ can be used in \eqref{eq:SIcancellation} to cancel a significant amount of SI. The residue of the SI can be canceled by the beamformers learned by the DRL algorithm.

\subsubsection{$\mathbf{H}_{AA}$ based SI cancellation} Finding an estimate of $\mathbf{H}_{AA}$ by maximum likelihood \cite{masmoudi2015maximum} or using deep learning \cite{muranov2021deep} is an active FD research area. If the communication system already has an estimate of $\mathbf{H}_{AA}^{}$ denoted by $\Tilde{\mathbf{H}}_{AA}^{}$, then the SI can be cancelled using \eqref{eq:SIcancellation} where
\begin{equation}
    \label{eq:noisyHAA}
    \hat{h} = \mathbf{w}_R^{}\Tilde{\mathbf{H}}_{AA}^{}\mathbf{w}_T^{}.
\end{equation}
We refer to this method as $\mathbf{H}_{AA}$-based SI-cancellation (HSIC).

The beamformers $\mathbf{w}_T^{}$ and $\mathbf{w}_R^{}$ in \eqref{eq:LSSIC} and \eqref{eq:noisyHAA} are predicted by the DRL agent at every time step. In the first stage, the LSSIC method helps to cancel the SI by estimating $\hat{h}$ by sending one pilot signal at every time step. On the other hand, if an estimate of $\mathbf{H}_{AA}^{}$ is already available at the BS, $\hat{h}$ can be estimated for SI-cancellation using this $\Tilde{\mathbf{H}}_{AA}^{}$. This does not require to send the pilot signal for estimating $\hat{h}$. The SI-canceled signal is then fed to the DRL agent for further learning in the second stage. Therefore, the quality of the beamformers learned by the DRL agent in the second stage depends on the SI-cancelled signal from the first stage. At the same time, with accurate learning of the beamformers, the SI-cancellation is also better. In the next subsection, we propose a DRL-based complete online solution that predicts RIS phases, BS active beamformers, and the transmit powers of BS and ULue. This method aims to cancel the residual SI that remains after the first stage by predicting beamformers without requiring CSI knowledge.

    \ifCLASSOPTIONtwocolumn
\begin{figure}[!t]
    \centering
    \includegraphics[scale=0.38]{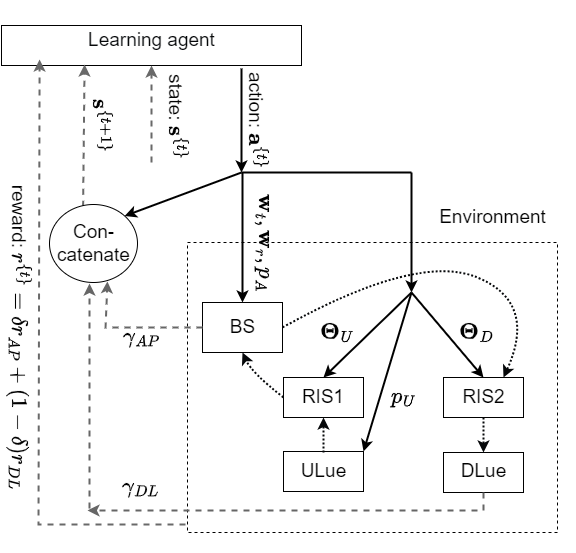}
    \caption{MDP formulation for RIS-based FD communication. The proposed algorithm can be deployed at the BS. The algorithm gives the action $\mathbf{a}^{\{t\}}$ (solid black) based on the state $\mathbf{s}^{\{t\}}$ (in dashed grey) generated at a previous time step. The environment reacts to these actions by sending the signal via RISs (dotted black) and returning the SINRs as the observations (in solid grey) indicate how good the actions are. Finally, the reward is calculated (dashed black) and fed as input to the learning agent. The SINR observations and actions at time step $t$ give the state for time step $(t+1)$. }
    \label{fig:irsdrl}
    \vspace{-3mm}
\end{figure}
\fi 

\subsection{Second stage: Learning-based Method}
Below, we model the problem of estimating phase shifts, beamformers, and transmit powers as a Markov decision process (MDP). An MDP has a state space $\mathcal{S}$, an action space $\mathcal{A}$, an initial distribution of space $p(\mathbf{s}^{\{1\}})$ and a stationary distribution for state transition that obeys Markov property i.e., $p(\mathbf{s}^{\{t+1\}}|\mathbf{s}^{\{t\}}, \mathbf{a}^{\{t\}})=p(\mathbf{s}^{\{t+1\}}|\mathbf{s}^{\{t\}}, \mathbf{a}^{\{t\}}, \dots, \mathbf{s}^{\{1\}}, \mathbf{a}^{\{1\}})$ and a reward function $r^{\{t\}}:\mathcal{S}\times\mathcal{A}\xrightarrow{}\mathbb{R}$. The overview of the MDP formulation for RIS-assisted FD communication is illustrated in Fig. \ref{fig:irsdrl}. The learning takes the state $\mathbf{s}^{\{t\}}$ as input. It predicts an action $\mathbf{a}^{\{t\}}$ that decides the transmit and receive beamforming vectors $\mathbf{w}_T^{\{t\}}$ and $\mathbf{w}_R^{\{t\}}$ at BS, the phase shift angles $\Bar{\Theta}_U^{\{t\}}$ and $\Bar{\Theta}_D^{\{t\}}$ at the two RISs and the transmit power at the BS $p_A^{\{t\}}$ and transmit power at the ULue $p_U^{\{t\}}$. The superscript $\{t\}$ denotes the quantities at time step $t$. The predicted action for time step $t$ is given by
\ifCLASSOPTIONtwocolumn
\begin{equation}
    \begin{aligned}
    \mathbf{a}^{\{t\}} = [&\Bar{\Theta}_U^{\{t\}}, \Bar{\Theta}_D^{\{t\}}, \text{real}(\mathbf{w}_T^{\{t\}}), \text{imag}(\mathbf{w}_T^{\{t\}}), \text{real}(\mathbf{w}_R^{\{t\}}), \\&\text{imag}(\mathbf{w}_R^{\{t\}}),  p^{\{t\}}_A, p^{\{t\}}_U].
\end{aligned}
\end{equation}
\fi
Using these predicted actions, the signal is transmitted from the ULue and the BS at time step $t$. The SINRs achieved at the BS and DLue depict the quality of the actions taken\footnote{We assume that the DRL agent acquires the information about DL SINR from ACK/NACK signals it receives from the DLue.}. The observation from the environment is the SINRs at both the BS and DLue denoted by $\gamma_{BS}^{\{t\}}$ and $\gamma_{DL}^{\{t\}}$ and are incorporated in the state $\mathbf{s}^{\{t+1\}}$ that goes inside the learning agent in the next iteration. The state for next time step $\mathbf{s}^{\{t+1\}}$ is represented by
\ifCLASSOPTIONtwocolumn
\begin{equation}
    \begin{aligned}
    \label{eq:newstate}
    s^{\{t+1\}}=[&\gamma^{\{t\}}_{BS}, \gamma^{\{t\}}_{DL}, \Bar{\Theta}_U^{\{t\}}, \Bar{\Theta}_D^{\{t\}}, \text{real}(\mathbf{w}_T^{\{t\}}), \text{imag}(\mathbf{w}_T^{\{t\}}), \\&\text{real}(\mathbf{w}_R^{\{t\}}), \text{imag}(\mathbf{w}_R^{\{t\}}),  p^{\{t\}}_A, p^{\{t\}}_U].
\end{aligned}
\end{equation}
\fi

The reward for time step $t$, given by $r^{\{t\}}$, is also provided as input to the learning agent along with the states. The existing literature uses the SNR as a reward \cite{feng2020deep}. However, in our work, the objective is to maximize the weighted sum of the UL and DL data rates given by $r_{BS}^{\{t\}} = \log_2 ( 1 + \gamma_{BS}^{\{t\}})$ and $r_{DL}^{\{t\}} = \log_2 ( 1 + \gamma_{DL}^{\{t\}} )$, respectively. So the reward at time step $t$ is taken as
\begin{equation}
    \label{eq:rewardR1}
    \qquad r^{\{t\}}=\delta r_{BS}^{\{t\}}+(1-\delta) r_{DL}^{\{t\}}.
\end{equation} 
Here, $r^{\{t\}}$ has the implicit knowledge of CSI that the agent can exploit by learning with the reward $r^{\{t\}}$.

The goal of the RL agent is to learn the policy $\bm{\pi}$ from the observations corresponding to each of the actions while maximizing the return $\mathbbm{r}^{\{t\}}(\gamma)$, where $\gamma\in[0,1]$ is the discounting factor. The policy maps the states and actions $\bm{\pi}:\mathcal{S}\xrightarrow{} \mathcal{A}$ at time step $t$ and obtains a reward $r^{\{t\}}(\mathbf{a}^{\{t\}},\mathbf{s}^{\{t\}})$. The return at time step $t$ is the discounted reward from time step $t$, and is given by $\mathbbm{r}^{\{t\}}(\gamma)=\sum_{t'=t}^\infty \gamma^{t'-t} r^{\{t\}}(\mathbf{a}^{\{t\}},\mathbf{s}^{\{t\}})$. 
The agent aims to find the policy $\bm{\pi}$ that maximizes the expected cumulative return $\mathbb{E}[\mathbbm{r}^{\{1\}}(\gamma)|\bm{\pi}]$. 

The learning agent calculates the value of a state $s$ by computing the expected return of state $s$ following a policy $\bm{\pi}$ given by $V^\pi(s)= \mathbb{E}\left[\mathbbm{r}^{\{1\}}(\gamma)| S_1 = s; \bm{\pi} \right]$. It denotes how good a state $s$ is for the agent to be in. The agent calculates the quality of action by using $Q$-function given by $Q^\pi(s, a) = \mathbb{E}\left[\mathbbm{r}_1(\gamma)|S_1 = s, A_1 = a; \bm{\pi} \right]$, which indicates how rewarding each action $a$ is when taken from a state $s$. \\The agent takes action at each step to maximize the $ Q$ value. The best policy $\bm{\pi}$ at any state $s$ can be chosen by a function approximation approach as the state-action pairs are continuous and can take infinite values\footnote{For the discrete state and actions, the tabular methods are used to calculate $Q^\pi(s, a)$.}. By learning the parameters of the function that operates on $(s, a)$, the agent can pick the policy $\bm{\pi}$ that gives the maximum return. The success of RL algorithms depends on the expressive power of the candidate functions, such as linear functions, radial basis functions, Fourier basis, neural networks, etc., that can maximize the return of policy $\bm{\pi}$. The DRL agent uses a policy approximated by a deep neural network that maps state $\mathbf{s}^{\{t\}}$ to action $\mathbf{a}^{\{t\}}$ for its extraordinary function approximation capabilities.

\subsection{Deep Reinforcement Learning based Predictor}
To train the DRL agent, an actor-critic method named deep deterministic policy gradient (DDPG) is used with continuous action space as used in several other works\cite{raj2022deep, feng2020deep,faisal2021deep}. The architecture of the deep network should be designed appropriately to obtain the predicted actions as the output of the architecture. Therefore, the vanilla DDPG method with the simple feed-forward network cannot be used. For a complete understanding, we discuss the DDPG algorithm briefly in the next paragraph before discussing the proposed actor architecture for predicting RIS phase shifts, beamformers, and transmit powers.

DDPG is a combination of $Q$-learning and policy gradient. The parameters of the policy (the deep neural network) are tuned so that the policy gives an output that improves the return. DDPG has two neural networks: an actor-network $\mathbbm{A}$ parameterized by $\bm{\omega}^a$, which predicts the action $\mathbf{a}^{\{t\}}$ based on the current state $\mathbf{s}^{\{t\}}$ and critic network $\mathbbm{C}$ parameterized by $\bm{\omega}^c$ which computes  $Q(\mathbf{s}^{\{t\}}, \mathbf{a}^{\{t\}})$ that is essentially the quality of the action taken by actor-network $\mathbbm{A}$ and hence encourages the actor-network to take \textit{better} actions through its feedback. Meanwhile, the critic network $\mathbbm{C}$ trains itself for better prediction by observing the rewards after each action. It first computes the Q-value of each action and then calculates the gradient of the error on its prediction of the Q-value of actions. DDPG maintains a replay buffer of finite size $\tau$ and samples the observations from the buffer in mini-batches to update the parameters to get stable, uncorrelated gradients for policy improvement. The buffer stays in the memory of the learning agent's computational device at the BS for efficient data storage, real-time learning, and flexibility. Storing experiences in memory allows quick access and updates, which is crucial for timely decision-making and continuous learning in DRL algorithms. Moreover, the learning agent is implemented at the BS, which can handle the required processing, aligning our solution with practical applications. Several DRL approaches have employed centralized training methodologies involving the actor-critic networks of a DDPG agent utilizing samples from the experience replay buffer in wireless applications \cite{taha2020deep,feng2020deep,lin2020deep,faisal2021deep}. The works mentioned above employ centralized training, where the agent and its associated replay buffer are assumed to be at the BS. DDPG solves large-dimension optimization problems comprising multiple users also \cite{huang2020reconfigurable}. Furthermore, DDPG can also solve more complex problems, such as hybrid beamforming in a multi-hop RIS-assisted network with multiple single-antenna users by centralized training \cite{huang2021multi}. These studies demonstrate the appropriateness of the DDPG algorithm in practical application scenarios.

DDPG also uses target networks with parameters $\bar{\bm{\omega}}^a$ and $\bar{\bm{\omega}}^c$ to avoid divergence in value estimation. The target network helps the learning agent to update the parameters of the active network based on the values from the target network, giving the learning agent a stable error value from which to learn. At each time step, the state $\mathbf{s}^{\{i\}}$ and the action taken $\mathbf{a}^{\{i\}}$ along with the reward obtained $r^{\{i\}}$ and the next state $\mathbf{s}^{\{i+1\}}$ is stored as an experience $(\mathbf{s}^{\{i\}}, \mathbf{a}^{\{i\}}, r^{\{i\}}, \mathbf{s}^{\{i+1\}})$ to the buffer $\mathcal{B}$. For training the actor and critic networks, $N$ samples are taken from $\mathcal{B}$, which is used to compute the gradients. For the critic network $\mathbbm{C}(\cdot|\bm{\omega}^c)$ to compute the Q-value for each state action-pair, an estimate of return for state $s^{\{i\}}$ in each sample is computed as
    \begin{align}
        y^{\{i\}} = r^{\{i\}} + \gamma \mathbbm{C}(\mathbf{s}^{\{i+1\}}, \mathbbm{A}(\mathbf{s}^{\{i+1\}}|
                                \bar{\bm{\omega}}^a)|\bar{\bm{\omega}}^c).
                                \label{eqn:critic_return}
    \end{align}
Once we observe a reward $r^{\{i\}}$ after taking an action $a^{\{i\}}$, based on the estimate for return, the mean squared Bellman error (MSBE) is computed as
    \begin{align}
        \mathcal{L} = \frac{1}{N} \sum \limits_{i} \left( 
                y^{\{i\}} - \mathbbm{C}(\mathbf{s}^{\{i\}}, \mathbf{a}^{\{i\}}|\bm{\omega}^c)
            \right)^2, \label{eqn:critic_msbe}
    \end{align}
    where, $\mathbbm{C}(\cdot)$ is the predicted output value of critic network with parameter $\bm{\omega}^c$ for the state $\mathbf{s}^{\{i\}}$ and action $\mathbf{a}^{\{i\}}$ before seeing the reward. Then, the critic network parameters are updated as
    \begin{align}
        \bm{\omega}^c \gets \bm{\omega}^c - \eta_c \nabla_{\bm{\omega}^c} \mathcal{L},
            \label{exp:critic_update}
    \end{align}
    where $\eta_c \ll 1$ is the stepsize for the stochastic update. For the actor-network, the update depends on both the gradient of action and the improvement in Q-value. The final update for updating parameters of the actor-network $\bm{\omega}^a$ is given by
    \begin{align}
        \bm{\omega}^a \gets \bm{\omega}^a + 
                            \eta_a \frac{1}{N} \sum \limits_{i} 
                            \left( \nabla_{\bm{\omega}^a} \mathbbm{A}(s) 
                                \nabla_a \mathbbm{C}(s,a) | _{a=\mathbbm{A}(s)} \right),
            \label{exp:actor_update}
    \end{align}
    where $\eta_a \ll 1$ is the update step size. Finally, the target network parameters are updated in every $U$ time step to provide stable value estimates using an exponentially weighted update as $\bar{\bm{\omega}}^c \gets \lambda \bm{\omega}^c+ (1 - \lambda) \bar{\bm{\omega}}^c$, and $\bar{\bm{\omega}}^a \gets \lambda \bm{\omega}^a+ (1 - \lambda) \bar{\bm{\omega}}^a$, with $\lambda \ll 1$. For a more detailed discussion on the DDPG algorithm, interested readers may refer to   \cite{silver2014deterministic,lillicrap2015continuous, raj2022deep}.

\subsection{Architecture of the  Proposed Neural Action Predictor}
\label{sec:actornetwork}
The DDPG algorithm deals with continuous action and continuous state space. The actor-network $\mathbbm{A}$ predicts action from the state space, and the role of critic network $\mathbbm{C}$ is to evaluate the Q-value of the actions. We propose a two-part neural architecture design to handle the huge action space of RIS phases, beamformers, and transmit powers in our problem. The first part is a feed-forward feature extractor (FE) with ReLU as an activation function and layer normalization. The output feature is then passed through the second part, consisting of multiple sub-networks (SN) to give the phase shifts, beamforming vectors, and transmit powers. The architecture of the proposed actor-network is illustrated in Fig. \ref{fig:actor}. With SN architectures, we can easily modify the architecture of the SNs using domain knowledge of the actions we want to predict. Later, in Sec. \ref{sec:quantgroup}, we show that one can change the $SN_1$ and $SN_2$ to predict discrete RIS phase shifts, keeping the other SNs unaltered. The critic network gives a Q-value of an action, which is a scalar value; therefore, we propose to use only a feed-forward neural network for the critic network $\mathbbm{C}$.
\ifCLASSOPTIONtwocolumn
\begin{figure}
    \centering
    \includegraphics[scale=0.55]{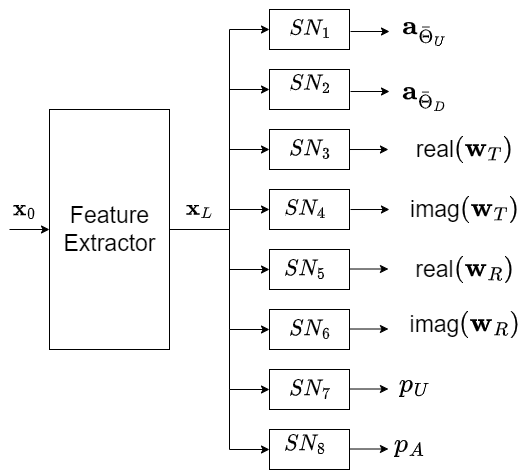}
    \caption{The proposed action predictor network.}
    \vspace{-3mm}
    \label{fig:actor}
\end{figure}
\fi

At time step $t$, the input to the feature extractor is the state $\mathbf{x}_0=\mathbf{s}^{\{t\}}$ that is of dimension $2+N_1^{}+N_2^{}+2M_t^{}+2M_r^{}+2$, where the first two entries are the observations $\gamma^{\{t-1\}}_{BS}, \gamma^{\{t-1\}}_{DL}$ from the previous time step $(t-1)$. Next, $N_1$ and $N_2$ entries are the phase shifts of RIS1 and RIS2, respectively, predicted at time step $(t-1)$. Next, $2M_t$ and $2M_r$ entries are the in-phase and quadrature values of the transmit and receive beamforming vectors, respectively. The final two entries are the transmit powers of ULue and BS. The proposed feature extractor has $L$ linear layers, each followed by layer normalization and ReLU activation. If the weights and biases of the $l^{th}$ layer are denoted by $\mathbf{W}_{l}^{}\in \mathbb{R}^{d_{l,o}^{}\times d_{l,i}^{}}$ and $ \mathbf{b}_{l}^{}\in \mathbb{R}^{d_{l,o}^{}}$, respectively, then the output of $l^{th}$ layer is given by $\mathbf{x}_l^{}=\text{ReLU}(\mathbf{W}_{l}^{}\mathbf{x}_{l-1}^{}+\mathbf{b}_{l}^{})$ for $l=1, \dots,L$ and $ \mathbf{x}_{l}\in \mathbb{R}^{d_{l,o}^{}}$. The output of the feature extractor is $\mathbf{x}_{L}$ and is input to each of the SNs, as shown in Fig. \ref{fig:actor}. The architectures of each SN are described below.

\subsubsection{RIS phases} The first two SNs that predict UL and DL RIS phases are parameterized by $(\mathbf{W}_{\Bar{\Theta}_U^{}}^{},\mathbf{b}_{\Bar{\Theta}_U^{}}^{})$ and $(\mathbf{W}_{\Bar{\Theta}_D^{}}^{},\mathbf{b}_{\Bar{\Theta}_D^{}}^{})$, respectively. The feature $\mathbf{x}_{L}^{}$ is passed via the SNs, and the outputs from them are
\ifCLASSOPTIONtwocolumn
\begin{equation}
    \begin{aligned}
        \mathbf{a}_{\Bar{\Theta}_U^{}}^{} &= \text{tanh}(\mathbf{W}_{\Bar{\Theta}_U^{}}^{} \mathbf{x}_{L}^{} + \mathbf{b}_{\Bar{\Theta}_U^{}}^{}) \text{ and } \\
        \mathbf{a}_{\Bar{\Theta}_D^{}}^{} &= \text{tanh}(\mathbf{W}_{\Bar{\Theta}_D^{}}^{} \mathbf{x}_{L}^{} + \mathbf{b}_{\Bar{\Theta}_D^{}}^{}),
            \label{eqn:actor_theta1_theta2}
    \end{aligned}
\end{equation}
\fi
respectively, where $\mathbf{W}_{\Bar{\Theta}_U^{}}^{}\in\mathbb{R}^{N_1^{}\times d_{L,o}^{}}$, $\mathbf{b}_{\Bar{\Theta}_U^{}}^{}\in\mathbb{R}^{N_1^{}}$, $\mathbf{W}_{\Bar{\Theta}_D^{}}^{}\in\mathbb{R}^{N_2^{}\times d_{L,o}^{}}$, and $\mathbf{b}_{\Bar{\Theta}_D^{}}^{}\in\mathbb{R}^{N_2^{}}$. Note that in most of the RL algorithms, the predicted continuous actions initially follow a Gaussian distribution with zero mean and standard deviation equal to one. So, the DRL algorithm cannot predict actions that are not symmetric. Therefore, while using a custom environment, it is necessary to first normalize the actions using tanh activation at the final layer of each SN and make them symmetric to take value in $[-1,+1]$ and then shift and scale to modify the actions as per the domain knowledge of RIS phases, beamformers, and transmit powers. For example, the phase shifts at RIS are found by shifting and scaling the corresponding actions to take values in the range $[0,2\pi]$ before using it in the environment. So, the RIS phase shifts are computed (in radian) as $\Bar{\Theta}_U^{}=2\pi\times(\tilde{\mathbf{a}}_{\Bar{\Theta}_U^{}}^{}+1)/2$ and $\Bar{\Theta}_D^{}=2\pi\times(\tilde{\mathbf{a}}_{\Bar{\Theta}_D^{}}^{}+1)/2$.

\subsubsection{Beamformers} The in-phase and quadrature parts of beamforming vectors can take any value between $[-1,+1]$; unlike the RIS phases, there is no need to shift and scale the actions. Therefore, the in-phase and quadrature parts of receive and transmit beamforming vectors are the outputs from the SNs, each of which has two fully connected layers followed by tanh activation. The SN for the in-phase component of the transmit-beamformer is parameterized by $(\mathbf{W}_{1, M_t^{}, I},\mathbf{b}_{1, M_t^{}, I},\mathbf{W}_{2, M_t^{}, I},\mathbf{b}_{2, M_t^{}, I})$. So, the action corresponding to the in-phase and quadrature components of the transmit beamforming vector is given by
\ifCLASSOPTIONtwocolumn
\begin{equation*}
    \begin{aligned}
        \mathbf{a}_{M_t^{},I}^{} &= \text{real}(\mathbf{w}_T^{}) =  \text{tanh}(\mathbf{W}_{2,M_t^{},I}^{}\\&\text{ReLU}(\mathbf{W}_{1,M_t^{},I}^{} \mathbf{x}_{L}^{} + \mathbf{b}_{1,M_t^{},I}^{})+\mathbf{b}_{2,M_t^{},I}^{}) \text{ and } \\
        \mathbf{a}_{M_t^{},Q}^{} &= \text{imag}(\mathbf{w}_T^{}) = \text{tanh}(\mathbf{W}_{2,M_t^{},Q}^{}\\&\text{ReLU}(\mathbf{W}_{1,M_t^{},Q}^{} \mathbf{x}_{L}^{} + \mathbf{b}_{1,M_t^{},Q}^{})+\mathbf{b}_{2,M_t^{},Q}^{}),
    \end{aligned}
\end{equation*}
\fi
where $\mathbf{W}_{1, M_t^{}, I}^{}$, $\mathbf{W}_{1, M_t^{}, Q}^{}\in\mathbb{R}^{M_t^{}\times b_{L,o}^{}}$, $\mathbf{W}_{2, M_t^{}, I}^{}$, $\mathbf{W}_{2, M_t^{}, Q}^{} \in \mathbb{R}^{M_t^{}\times M_t^{}}$, and $\mathbf{b}_{1, M_t^{}, I}^{}$, $\mathbf{b}_{2, M_t^{}, I}^{}$, $\mathbf{b}_{1, M_t^{}, Q}^{}$, $\mathbf{b}_{2, M_t^{},Q}^{} \in \mathbb{R}^{M_t^{}}$. A similar expression can be written for the in-phase and quadrature components of receive beamforming vectors. The output beamformers are then normalized to have a unit norm.

\subsubsection{Transmit powers} The next two SNs predict the transmit powers of ULue and the BS. Each of the two SNs consists of a dense layer followed by tanh activation. The output of these two SNs parameterized by $(\mathbf{w}_{p_U}^{},b_{p_U}^{})$, $(\mathbf{w}_{p_A}^{},b_{p_A}^{})$ are given by 
\ifCLASSOPTIONtwocolumn
\begin{equation}
    \begin{aligned}
    a_{p_U}^{} &= \text{tanh}(\mathbf{w}_{p_U}^{} \mathbf{x}_{L}^{} + b_{p_U}^{}), \,\,
    a_{p_A}^{} = \text{tanh}(\mathbf{w}_{p_A}^{} \mathbf{x}_{L}^{} + b_{p_A}^{}),
    \label{eqn:actor_transmitpower}
\end{aligned}
\end{equation}
\fi
where $\mathbf{w}_{p_U}, \mathbf{w}_{p_A}^{}\in\mathbb{R}^{1\times b_{L,o}^{}}$ and $b_{p_U}^{}, b_{p_A}^{}\in\mathbb{R}$. The output from SN, which is in the range $[-1,+1]$, is shifted and scaled to the range $[0,P_u^{max}]$ and $[0,P_a^{max}]$ before using them in the environment 
\ifCLASSOPTIONtwocolumn
\begin{equation}
 \begin{aligned}
     p_a^{}&=(a_{p_A^{}}+1)/2\times P_a^{max}, \,\,
     p_u^{}=(a_{p_U^{}}+1)/2\times P_u^{max}.
 \end{aligned}
\end{equation}
\fi
Therefore, the final action taken on the environment is the vector $\mathbf{a}^{\{t\}}=[\Bar{\Theta}_U$, $\Bar{\Theta}_D $, $ \mathbf{a}_{M_t, I} $, $ \mathbf{a}_{M_t, Q}$, $\mathbf{a}_{M_r, I}$, $\mathbf{a}_{M_r, Q} $, $ p_a $, $ p_u]$. So, the dimension of the output layer of the actor-network is $N_1^{}+N_2^{}+4M_{t}^{}+2$. The critic network gives the Q-value of the state-action pair, which is just a scalar. Therefore, the dimension of the output layer of the critic network is one.

\textbf{Exploration of action space gives faster convergence:} For better convergence, usually in DRL applications\cite{lillicrap2015continuous}, the Ornstein Uhlenbeck\footnote{\url{https://stable-baselines.readthedocs.io/en/master/modules/ddpg.html\#stable_baselines.ddpg.OrnsteinUhlenbeckActionNoise}}(OU) action noise is added to the outputs from the actor-network to find an action. We propose adding Gaussian noise to the action space for a faster convergence than OU. Before any shifting and scaling of the actions, the action space of both phase shifts and beamformers is explored by adding a Gaussian noise $\mathcal{N}(0,\sigma_{expl})$ to both phase shifts and beamforming vectors. The standard deviation of noise $\sigma_{expl}$ is not very high and slowly decreases linearly over time with zero standard deviation at the end of training. The modified output from the SNs corresponding to the phase shifts and the in-phase and quadrature components of transmit and receive beamformers can be written as
\ifCLASSOPTIONtwocolumn
\begin{equation}
    \begin{aligned}
        \tilde{\mathbf{a}}_{\Bar{\Theta}_U^{}}^{} &= \mathbf{a}_{\Bar{\Theta}_U^{}}^{} + \mathbf{n}_1^{}, 
        \tilde{\mathbf{a}}_{\Bar{\Theta}_D^{}}^{} = \mathbf{a}_{\Bar{\Theta}_D^{}}^{} + \mathbf{n}_2^{}, \\ 
        \tilde{\mathbf{a}}_{M_t^{},I}^{} &= \mathbf{a}_{M_t^{},I}^{} + \mathbf{n}_3^{}, 
        \tilde{\mathbf{a}}_{M_t^{},Q}^{} = \mathbf{a}_{M_t^{},Q}^{} + \mathbf{n}_4^{}, \\
        \tilde{\mathbf{a}}_{M_r^{},I}^{} &= \mathbf{a}_{M_r^{},I}^{} + \mathbf{n}_5^{}, 
        \tilde{\mathbf{a}}_{M_r^{},Q}^{} = \mathbf{a}_{M_r^{},Q}^{} + \mathbf{n}_6^{},
        \label{eqn:actor_beamformers}
    \end{aligned}
\end{equation}
\fi
where $\mathbf{n}_1^{}, \mathbf{n}_2^{}, \mathbf{n}_3^{}, \mathbf{n}_4^{}, \mathbf{n}_5^{}, \mathbf{n}_6^{} \sim \mathcal{N}(0,\sigma_{expl})$.

\textbf{Initialization of the network parameters:} The kernel weight matrices corresponding to all the layers except the last layer of all the SNs are initialized with the Glorot uniform initializer \cite{glorot2010understanding}. If one also initializes the last layers with the Glorot uniform initializer, the predicted actions may take values near the two extreme ranges of the action space. Therefore, for better convergence, the last layers of all the SNs of the Neural Action Predictor network are initialized with a random uniform initializer with comparatively smaller values between $[-3\times10^{-3},3\times10^{-3}]$. 

The proposed DRL method, together with the proposed least square-based SI-cancellation (LSSIC) method, predicts the action that consists of RIS phases, the beamformers, and transmit powers. This does not assume any knowledge of CSI; therefore, the signaling overhead due to channel estimation can be avoided. The only feedback required from the environment for the agent is the SINR at DLue. Because of such minimal signaling feedback (MSF) to the agent, the proposed method is referred to as the \textit{``MSF-DRL"} method. The proposed method is demonstrated in Alg. \ref{alg:proposed_irsdrl}.

\ifCLASSOPTIONtwocolumn
\begin{algorithm}[!t] 
    \caption{Proposed algorithm MSF-DRL}
    \begin{algorithmic}[1]
        \State \textbf{Parameters:} Set $\gamma$, $\tau$, $M$, $\lambda$, $U$, $\eta_a$ and $\eta_c$. Initialize $\mathbbm{A}(s|\bm{\omega}^a)$ and $\mathbbm{C}(s,a|\bm{\omega}^c)$ with random weights $\bm{\omega}^a$ and $\bm{\omega}^c$ respectively. Initialize target networks with weights as $\bar{\bm{\omega}}^a \gets 
                \bm{\omega}^a$ and $\bar{\bm{\omega}}^c \gets \bm{\omega}^c$. Create an empty replay buffer $\mathcal{B} \gets \{\}$ with size $\tau$.
        \For {episode $= 1 \ldots M$}
            \State Select a random valid action $\mathbf{a}^{\{0\}}$.
            \State Observe $\gamma_{BS}^{\{0\}}$ and $\gamma_{DL}^{\{0\}}$ and get state $s^{\{1\}}$ from \eqref{eq:newstate}. 
            \For {$t = 1 \ldots T$}
                \State Set $\mathbf{a}_{\Bar{\Theta}_U}, \mathbf{a}_{\Bar{\Theta}_D}, a_{p_U}, a_{p_A}$, $\tilde{\mathbf{a}}_{M_r}$, and $\tilde{\mathbf{a}}_{M_t}$ according to \eqref{eqn:actor_theta1_theta2}, \eqref{eqn:actor_transmitpower}, and  \eqref{eqn:actor_beamformers}.
                \State Get new state observation $\mathbf{s}^{\{t+1\}}$ from \eqref{eq:newstate}.
                \State Compute reward $r^{\{t\}}$ from \eqref{eq:rewardR1}.
                
                \State Update replay buffer with experience as $\mathcal{B} \gets \mathcal{B} \cup (\mathbf{s}^{\{t\}}, \mathbf{a}^{\{t\}}, r^{\{t\}}, \mathbf{s}^{\{t+1\}})$.
                \If {$|\mathcal{B}| \geq \tau$}
                    \State Delete oldest experience from $\mathcal{B}$.
                \EndIf
                \State Sample $N$ experiences $(\mathbf{s}^{\{i\}}, \mathbf{a}^{\{i\}}, r^{\{i\}}, \mathbf{s}^{\{i+1\}})$ from $\mathcal{B}$.
                \State Compute return for each experience $y_i$.
                \State Compute Mean Square Bellman Error as $\mathcal{L}$.
                \State Update $(\bm{\omega}^c,\bm{\omega}^a)$ and $(\bar{\bm{\omega}^c}, \bar{\bm{\omega}^a})$.
            \EndFor
        \EndFor
    \end{algorithmic}
    \label{alg:proposed_irsdrl}   
\end{algorithm}
\fi

Although the MSF-DRL method has minimum signaling feedback to the BS, the BS still needs to send all real-valued RIS phases to the passive elements of the RISs, typically via a dedicated wired or wireless channel. Therefore, to reduce the signaling from BS to RISs, a quantized MSF-DRL method is designed to select only RIS phases from a limited number of possible phase values in the next section. Picking the quantized phase value is not easy for a traditional algorithm as it may produce a suboptimal solution\cite{wu2019beamforming,shekhar2022instantaneous,subhash2022max}. By changing the architecture slightly, we enable our proposed DRL agent to learn the quantized RIS phases. Moreover, quantized phases reduce the action space to a great extent, thus achieving a better convergence. We also attempt to further reduce the signaling by using a simple grouping strategy for the RIS phase. 

\section{Proposed Quantized and Grouped Quantized MSF-DRL}
\label{sec:quantgroup}
One of the major issues with the RIS is that the phase shifts need to be communicated from the agent placed at the BS, along with the message signal at certain intervals. This is a signaling overhead and ideally should be as minimum as possible. We leverage the DRL framework to learn quantized RIS phase shifts to reduce the signaling overhead and call it MSF(Q)-DRL. Instead of real values (i.e., $64$-bit), every element of the RISs is represented with just $n$ bits ($n\ll64$), so that instead of a $64$-bit value, only $n$-bit information is transmitted from the BS to the RIS. Thus, the number of possible phase shifts that each of the passive elements can provide is $Q=2^{n}$ and which are given by $\mathbf{p}=2\pi/2^n\times[0,\dots,2^n-1]$ radians. The architecture of the first two SNs for predicting RIS phases is now modified so that instead of a single phase value, they predict the probabilities of picking a phase value out of the possible values in $\mathbf{p}$. So, one out of $2^{n}$ elements will have a high probability for a particular RIS element. The SN for predicting the phase shift of RIS1 takes the feature $\mathbf{x}_L$ as input and passes it through $N_1$ fully connected layers for $N_1$ elements, each with $2^{n}$ neurons, followed by a softmax activation on each of the $N_1$ outputs. The output of the first SN is given by $\mathbf{a}_{i,\Bar{\Theta}_U} = \text{softmax}(\mathbf{W}_{i,\Bar{\Theta}_U} \mathbf{x}_{L} + \mathbf{b}_{i,\Bar{\Theta}_U}); \hspace{1mm} \forall i \in \{1,\dots,N_1\},$ where $\mathbf{W}_{i,\Bar{\Theta}_U}\in\mathbb{R}^{2^n\times d_{L,o}}$, $\mathbf{b}_{\Bar{\Theta}_U}\in\mathbb{R}^{2^n}$. Therefore, the RIS phase shift of $i^{th}$ element of RIS1 is given by
\begin{equation}
    \Bar{\Theta}_{i,U}= p[\argmax{\mathbf{a}_{i,\Bar{\Theta}_U}}]\hspace{1mm} \forall i \in \{1,\dots,N_1\}. 
\end{equation}
The RIS phase shifts of $N_1$ elements are concatenated together to form the output of SN1, i.e., $\Bar{\Theta}_U^{} = [\Bar{\Theta}_{1,U}, \Bar{\Theta}_{2,U}, \dots, \Bar{\Theta}_{N_1,U}]$. The same applies to predicting RIS2 phases, too. The architectures of all the other SNs are the same as those of the previous method. The proposed algorithm can handle a pseudo-discrete action space where RIS phase shifts are discrete and beamformers and transmit powers are continuous. Deep Q-networks (DQN) \cite{mnih2013playing} are suitable for only discrete action space and hence cannot be used for a mixed action space like ours. By using MSF(Q)-DRL, we have reduced the dimension of the action space and the signaling overhead to the two RISs.

A set of spatially close passive elements are grouped to further reduce the signaling. Then, instead of $N$ phase shifts, $M~(M<N)$ phases could be fed to the RISs to reduce the overhead, though this is not optimal. Multiple spatially close elements can be grouped to have the same phase shifts. These grouped phase MSF(Q)-DRL methods are referred to as GP(M)-MSF(Q)-DRL. While conventional optimization algorithms will find this difficult, the proposed DRL agent can learn these $M$ phase shifts directly. It will not require a completely new DRL architecture, and only a slight modification to the existing architecture will suffice.

\section{Numerical Study and Discussion}
In this section, we provide numerical results to investigate the performance of the proposed methods quantitatively. The alignment of RISs, users, BS, and the channel parameters closely follow \cite{cai2021intelligent}. For the simulation setup, consider that the BS is situated at the center $(0,0)$, and there is no direct path from the BS to the UL and DL users. Two RISs are incorporated and placed at $(50,22)$ and $(50,-22)$ to promote communication without a direct path. The two RISs are parallel to the xz plane, and the distance between them is $44$m, and the ULue and DLue are static at $(50,20)$ and $(50,-20)$, respectively, as shown in Fig.\ref{fig:metrocity}. The maximum transmit power allowable at ULue and BS are $p_U^{max}=50$mW and $p_A^{max}=1$W, respectively \cite{peng2021multiuser,pan2020multicell}. For the moving UE scenario, the channels are evolved by Markov Chain Monte Carlo simulation, where the samples are drawn after updating the UE location after each timestep following a mobility model. The UEs move in a square area of $100m^2$ with an average speed of $1m$ per time step \cite{evmorfos2022deep}. At the beginning of every episode, the positions of the UEs are initialized randomly, as every episode in the DRL setup is an independent game.

The BS-RIS and the RIS-user channels $\mathbf{F}_{AI}^{}$, $\mathbf{G}_{IA}^{}$, $\mathbf{g}_{DI}^{}$ and $\mathbf{f}_{IU}^{}$ are considered to have LOS. They are distributed as Rician with the Rician $K$-factors $\beta_{IA}^{}$ and $\beta_{UI}^{}$ for the channels between the BS and RIS and between the RIS and users, respectively \cite{pan2020multicell, peng2021multiuser, feng2020deep, cai2021intelligent}. The channel between RIS2 and DLue is given by
\begin{align}
\mathbf{g}_{DI}^{}&=\sqrt{\frac{\beta_{UI}^{}}{1+\beta_{UI}^{}}}\mathbf{g}_{DI}^{LOS} + \sqrt{\frac{1}{1+\beta_{UI}^{}}}\mathbf{g}_{DI}^{NLOS}.
\end{align}
The deterministic LOS component can be modeled and predicted with relative certainty based on the geometry and physical characteristics of the transmission environment, such as the LOS angles. Therefore, LOS components are found by using the transmit and receive beam steering vectors $\mathbf{a}_t$ and $\mathbf{a}_r$. For example, the LOS component of the channel $\mathbf{g}_{DI}$ can be written as $\mathbf{g}_{DI}^{LoS} = \mathbf{a}_r^{}(\theta_{AOA}^{}, \phi_{AOA}^{})\mathbf{a}_t^H(\theta_{AOD}^{}, \phi_{AOD}^{})$, where $\theta_{AOA}^{}$ and $\phi_{AOA}^{}$ are the elevation and azimuth angles of arrival at the receiver antenna array at BS whereas $\theta_{AOD}^{}$ and $\phi_{AOD}^{}$ are the elevation and azimuth angles of departure from the transmit antenna array at BS. We measure $\theta$ from the +z-axis and $\phi$ from the +x-axis. For the calculations of the steering vectors, the ULA antenna array of the BS is assumed to be on the y-axis and the RISs on the xz plane. The receive beam steering vector $\mathbf{a}_r^{}(\theta_{AOA}^{}, \phi_{AOA}^{}) = 1$ as the DLue is omnidirectional and transmit beam steering vector $\mathbf{a}_t^{}(\theta_{AOD}^{}, \phi_{AOD}^{}) = \mathbf{a}_{N_{2z}^{}}^{}(\theta_{AOD}^{})\otimes \mathbf{a}_{N_{2x}^{}}^{}(\theta_{AOD}^{}, \phi_{AOD}^{})$ where 
\ifCLASSOPTIONtwocolumn
\begin{equation}
\begin{aligned}
\label{eq:aoaaod}
    \mathbf{a}_{N_{2z}^{}}^{}(\theta_{AOD}^{}) &= [1, \dots, e^{j2\pi \frac{d_a}{\lambda} n_{2z}^{} \cos{\theta_{AOD}^{}}}, \dots, \\& e^{j2\pi \frac{d_a}{\lambda} (N_{2z}^{}-1) \cos{\theta_{AOD}^{}}}]^H\\
   \mathbf{a}_{N_{2x}^{}}^{}(\theta_{AOD}^{}, \phi_{AOD}^{})  &= [1, \dots, e^{j2\pi \frac{d_a}{\lambda} n_{2x}^{} \sin{\theta_{AOD}^{}}\cos{\phi_{AOD}^{}}}, \dots, \\& e^{j2\pi \frac{d_a}{\lambda} (N_{2x}^{}-1) \sin{\theta_{AOD}^{}}\cos{\phi_{AOD}^{}}}]^H,
\end{aligned}
\end{equation}
\fi
and $d_a$ is RIS element separation distance. The channels for other Rician-distributed links can be expressed similarly. For more details on the linear and planar array steering vectors, readers are requested to refer to Section 6.6.2 and Section 6.10.1 of \cite{balanis2015antenna}. The NLOS components are uncertain due to scattering or multi-path effects and are generated as time-dependent Rayleigh fading using the Jakes' method with a Doppler frequency of $100$Hz \cite{jakes1994microwave}. The path loss between two points with distance $d$ is modeled as \cite{5gmodel2016},
\begin{equation}
    PL(f_c,d)_{dB} = -20\log_{10}(4\pi f_c/c) - 10\alpha\log(d/D_0),
\end{equation}
where $f_c$ is the carrier frequency, $D_0=1$m, $\alpha$ is the path loss exponent. The path loss components in the BS-RIS and RIS-user links are $\alpha_{AI}=\alpha_{IU}=2.2$. The direct paths from the BS to the users are blocked. The users are considered to be in an urban region; therefore, the channels are considered Rayleigh distributed with higher path loss coefficients $\alpha_{AU}=3.35$ \cite{rappaport2010wireless}. The inter-RIS interference is considered with the channel $\mathbf{f}_{DI}^{}$ between RIS1 and DLue and with channel $\mathbf{g}_{IU}^{}$ between RIS2 and ULue, which are considered to be Rayleigh distributed with a path loss coefficient of $\alpha_{R}=3.35$. We consider inter-user interference with the channel $g$ between the ULue and the DLue. Assuming that the channel between ULue and DLue is shadowed, we treat $g$ also as Rayleigh distributed with a path loss coefficient of $\alpha_{U}=4.5$. We model the channel matrix between transmit and receive arrays at BS, denoted by $\mathbf{H}_{AA}$ with Rayleigh ﬂat fading \cite{riihonen2011hybrid}. Since each implementation of a particular analog/digital SI cancellation scheme can be characterized by a speciﬁc residual power, the elements of $\mathbf{H}_{AA}$ can be modeled as i.i.d. $\mathcal{CN}(0,\sigma_{AA}^2)$ random variables (RVs) \cite{mohammadi2015full}.

The hyperparameter $\delta$ is set as $0.5$; hence, the UL and DL data rates are given equal weightage. The Rician $K$-factors are $\beta_{IA}=9$dB and $\beta_{UI}=6$dB \cite{cai2021intelligent}. The bandwidth where the system operates is $100$MHz \cite{taha2020deep}, the noise power density is $-174$dBm/Hz \cite{peng2021multiuser}, and the carrier frequency $f_c$ is $3.5$GHz \cite{taha2020deep}. For the DRL agent, the discounting factor $\gamma=0.6$, buffer size $\tau=10000$, and the learning rates of actor and critic networks are $0.0001$ and $0.001$, respectively \cite{raj2022deep}. The experiment is run for an initial $100$ episode (unless otherwise specified), and each episode has $1000$ time steps. The results are averaged over $8$ independent runs. The experiments are performed on an NVIDIA GeForce RTX 2080 Ti GPU. The benchmark metrics used for studying the performance are UL and DL data rates with the unit bits/sec/Hz. 

Below, we compare the performance between the proposed and baseline competitor methods.

\subsubsection{RandPSBF} The RandPSBF agent receives no SINR feedback from the environment. It tries to randomly predict the RIS phases and beamformers without any input/feedback to the agent. In our simulations, the RandPSBF agent predicts the RIS phases and the beamformers randomly from $\mathcal{U}(0,2\pi)$ and $\mathcal{U}(-1,1)$, respectively. The transmit powers of ULue and BS are set to be the maximum allowable transmit powers of ULue and BS, denoted as $p_U^{max}$ and $p_A^{max}$, respectively.

\subsubsection{OUPSBF} The agent OUPSBF makes use of our proposed DRL framework as discussed in Sec. \ref{sec:actornetwork}, except for the action-noise, where the agent adds the Ornstein Uhlenbeck (OU) noise to the RIS phases and beamformer, which is a standard exploration strategy for DRL. Therefore, $\bm{n}_1, \bm{n}_2, \bm{n}_3, \bm{n}_4, \bm{n}_5,$ and $ \bm{n}_6$ in \eqref{eqn:actor_beamformers} are all generated using OU process.

\subsubsection{Proposed MSF-DRL-LSSIC and MSF-DRL-HSIC} The proposed two-stage learning algorithm provides a complete solution for RIS phase shift, active beamformers at the BS, and transmit powers at BS and ULue when the CSI and residual SI knowledge are unavailable. It can learn the beamformers successfully even when the residual SI power is as high as $\sigma_{AA}^2=0.1$. If an estimate (even if slightly noisy) of residual SI is available, MSF-DRL-HSIC can be used. We assume that an estimate of $\mathbf{H}_{AA}$ with a noise variance of $10^{-12}$ is available to the MSF-DRL-HSIC agent. The critic network and the feature extractor actor network for LSSIC and HSIC are feed-forward networks with two layers, each with $100$ neurons. At the beginning of every episode, the MSF-DRL agent chooses actions randomly - phase shifts from a uniform random distribution $\mathcal{U}(0,2\pi)$; in-phase and quadrature components of the beamformers from $\mathcal{U}(-1,1)$. Note that, at the beginning of every episode, if the agent chooses actions from a normal distribution instead of uniform, the convergence is much slower, and the final solution sometimes converges to a sub-optimal point. Initializing actions with uniform distribution allows the agent to explore the action space evenly, and therefore, the convergence is better. The transmit powers of the ULue and BS are initialized with values from $\mathcal{U}(p_U^{max}/3,2p_U^{max}/3)$ and $\mathcal{U}(p_A^{max}/3,2p_A^{max}/3)$, because the convergence may suffer if transmit powers are initialized close to zero or maximum transmit power. MSF-DRL uses a Gaussian action noise with zero mean and linearly decaying standard deviation (SD) with an initial SD of $0.3$ decaying over $100$ episodes.  

\subsubsection{PerfCSI-DRL and NoisCSI-DRL} To appreciate the power of the proposed method, we also perform a hypothetical experiment called ``PerfCSI-DRL" to predict only the RIS phases where \textit{the perfect CSI knowledge, including the residual SI, is available} to the agent, therefore, serves as a benchmark. Here, the agent calculates the beamformers based on the Zero Forcing (ZF) and minimum mean square error (MMSE) principle \cite{suraweera2014low,mohammadi2015full,faisal2021deep} that needs perfect CSI. 
Here, we consider MRC-RX/ZF-TX scheme, where the BS employs maximum ratio combining (MRC) processing for the received signal and ZF processing for the transmitted signal that makes use of the transmitting antennas to completely cancel the SI by using transmitter diversity\footnote{If the receiving antennas are used to cancel the residual SI completely, then the BS may employ ZF processing for the received signal and maximum ratio transmission (MRT) processing for the transmitted signal, resulting in ZF-RX/MRT-TX scheme. MRC-RX/ZF-TX and ZF-RX/MRT-TX can work when the system has higher residual SI. However, when the residual SI is low, MRC processing at the receiver and MRT processing at the transmitter (MRC-RX/MRT-TX) perform better. The MRC-RX/MRT-TX scheme does not need information about $\mathbf{H}_{AA}$.}. In the presence of high residual SI, ZF processing completely cancels the SI by using the transmitting antennas with $M_t>1$. The optimal ZF precoder maximizes the achievable UL and DL sum rate and is the solution of 
    \begin{equation}
    \begin{aligned}
        \mathcal{P}_2: \quad  \max_{\mathbf{w}_T} \hspace{2mm} & r_{DL}^{} + r_{BS}^{} \\  \text{s.t.} \hspace{2mm} & ||\mathbf{w}_R^{MRC}\mathbf{H}_{AA}^{}\mathbf{w}_T^{}||^2=0, \quad ||\mathbf{w}_T^{}||^2=1,
        \label{eq:MRCZF}
    \end{aligned}
    \end{equation}
    where
    \begin{equation}
        \begin{aligned}
        \label{eq:mrczfrx}
        \mathbf{w}^{MRC}_r = \frac{(\mathbf{h}_{AU}^{}+\mathbf{F}_{AI}^{}\mathbf{\Theta}_U^{}\mathbf{f}_{IU}^{}+\mathbf{G}_{IA}^{T}\mathbf{\Theta}_D^{}\mathbf{g}_{IU}^{})^H}{||\mathbf{h}_{AU}^{}+\mathbf{F}_{AI}^{}\mathbf{\Theta}_U^{}\mathbf{f}_{IU}^{}+\mathbf{G}_{IA}^{T}\mathbf{\Theta}_D^{}\mathbf{g}_{IU}^{}||}.
        \end{aligned}
    \end{equation}
After substituting the value of $r_{DL}^{}$ and $r_{BS}^{}$ in \eqref{eq:MRCZF} and simplifying, the  objective function is given by
\begin{equation}
    \begin{aligned}
    \label{eq:optimizationmrczf}
        \max_{\mathbf{w}_T^{}} \hspace{2mm}& |(\mathbf{h}_{DA}^{}+\mathbf{g}_{DI}^{}\mathbf{\Theta}_D^{}\mathbf{G}_{IA}^{}+\mathbf{f}_{DI}^{}\mathbf{\Theta}_U^{}\mathbf{F}_{AI}^{T})\mathbf{w}_T^{}|^2 \\  \text{s.t.} \hspace{2mm} & ||\mathbf{w}_R^{MRC}\mathbf{H}_{AA}^{}\mathbf{w}_T^{}||^2=0,  \quad ||\mathbf{w}_T^{}||^2=1.
    \end{aligned}
\end{equation}
Following the solution of a similar optimization problem in \cite{mohammadi2015full}, the precoder $\mathbf{w}_T$ is in the orthogonal complement space of $\mathbf{w}_R^{MRC}\mathbf{H}_{AA}^{}$. The orthogonal projection onto the orthogonal complement of the column space of $\mathbf{w}_R^{MRC}\mathbf{H}_{AA}^{}$ is given by 
\begin{equation*}
    \begin{aligned}
    \Pi^{\perp}_{\mathbf{H}_{AA}^{\dagger}\mathbf{w}_R^{MRC\dagger }}  &= \mathbf{I}_{M_t^{}}^{} \\ - \mathbf{H}_{AA}^{\dagger}\mathbf{w}_R^{MRC\dagger}& (\mathbf{w}_R^{MRC}\mathbf{H}_{AA}\mathbf{H}_{AA}^{\dagger}\mathbf{w}_R^{ MRC\dagger})^{-1}\mathbf{w}_R^{MRC}\mathbf{H}_{AA}^{}.
\end{aligned}
\end{equation*}
The optimal solution of \eqref{eq:optimizationmrczf} is given by 
\begin{align}
\label{eq:mrczftx}
    \mathbf{w}_T^{ZF} = \frac{\Pi^{\perp}_{\mathbf{H}_{AA}^{\dagger}\mathbf{w}_R^{MRC\dagger }}(\mathbf{h}_{DA}^{}+ \mathbf{g}_{DI}^{}\mathbf{\Theta}_D^{}\mathbf{G}_{IA}^{}+\mathbf{f}_{DI}^{}\mathbf{\Theta}_U^{}\mathbf{F}_{AI}^{T})^{\dagger}}{||\Pi^{\perp}_{\mathbf{H}_{AA}^{\dagger}\mathbf{w}_R^{MRC\dagger }}(\mathbf{h}_{DA}^{}+    \mathbf{g}_{DI}^{}\mathbf{\Theta}_D^{}\mathbf{G}_{IA}^{}+\mathbf{f}_{DI}^{}\mathbf{\Theta}_U^{}\mathbf{F}_{AI}^{T})^{\dagger}||}.
\end{align}
where $\dagger$ represents conjugate transpose.
The PerfCSI-DRL method can be called semi-oracle as it needs to periodically receive the CSI to calculate the beamformers $\mathbf{w}^{MRC}_R$ and $\mathbf{w}_T^{ZF}$. The overhead for CSI update can be big matrices for nodes with multiple antennae, whereas our proposed method utilizes only the received DL SINR, which is a single scalar value signaled from the DLue. The UL SINR is already present at the BS, so it does not need to be signaled. Different channel estimation methods can be used to estimate the CSI, which may not be exact. In practical scenarios, the CSI available to the agent is noisy, including a noisy estimate of residual SI channel $\mathbf{H}_{AA}$. The resulting method is referred to as the ``NoisCSI-DRL" method.

\subsubsection{Conventional-FD-DRL} The usual DRL algorithms that find RIS phase shifts calculate the beamformers using the MRT approach for HD and the semi-closed optimal solution approach for FD setup \cite{faisal2021deep}. Here, the system model differs from ours, with one receiver antenna, user, and BS. They use perfect knowledge of CSI and a residual SI in the range $\sigma_{AA}^2<10^{-9}$ \cite{faisal2021deep}.

\ifCLASSOPTIONtwocolumn
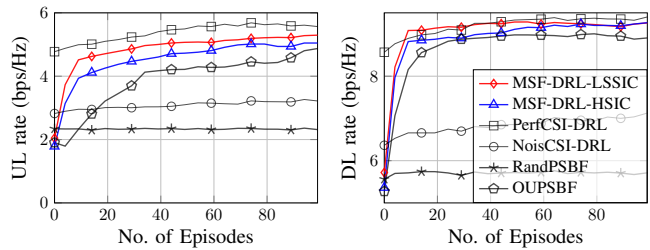
\begin{figure}[!t]
\centering
\pgfplotstableread[col sep = comma]{./datainterris/data_01UEUL_01UEDL_BS08_rateEvo_wsic_staticUE_wexpltnBF_LSSIC_urban_lowpathloss.csv}\datatableone
\tikzstyle{mark_style} = []

\pgfplotstableread[col sep = comma]{./datainterris/data_01UEUL_01UEDL_BS08_rateEvo_wsic_staticUE_FixedBFMRCZF_urban_lowpathloss_BS10_RIS36.csv}\datatablefive

\pgfplotstableread[col sep = comma]{./datainterris/data_01UEUL_01UEDL_BS08_rateEvo_wsic_staticUE_FixedBFMRCZF_noisyCSIpercent100_urban_lowpathloss_BS10_RIS36.csv}\datatableeleven

\pgfplotstableread[col sep = comma]{./datainterris/data_01UEUL_01UEDL_BS08_rateEvo_wsic_staticUE_wexpltnBF_LSSIC_urban_lowpathloss_randompsbf_B10R36.csv}\datatableweightedsumrandompsbf

\pgfplotstableread[col sep = comma]{./datainterris/data_01UEUL_01UEDL_BS08_rateEvo_wsic_staticUE_wexpltnBF_LSSIC_urban_lowpathloss_OU_B10R36.csv}\datatablewexpOU

\pgfplotstableread[col sep = comma]{./datainterris/data_01UEUL_01UEDL_BS08_rateEvo_wsic_staticUE_wexpltnBF_LSSIC_urban_lowpathloss_NoisyHAA.csv}\datatableoneirsNHAAsicstat
\tikzstyle{mark_style} = []

\begin{subfigure}[b]{0.495\linewidth}
\centering
\resizebox{\linewidth}{!}{
\begin{tikzpicture}[thick,scale=0.99]
    \begin{axis}[
        width=7cm,
        height=5.5cm,
        xmin=0,
        xmax=99,
        ymin=0,
        ymax=6,
        grid=major,
        xlabel={No. of Episodes},
        ylabel={UL rate (bps/Hz)},
        xlabel style={at={(0.50,0.03)}},
        ylabel style={at={(0.1,0.50)}},
        label style={font=\large},
        legend cell align={left},
        legend style={fill opacity=0.7, draw opacity=1.0, text opacity=1.0, font=\small}
        ]
                   
        \addplot[red, solid, thick, 
                mark=diamond, mark size={2.5}, every mark/.append style={solid, fill=gray}, mark repeat=3, opacity=0.9
                ] 
            table [x=x_data, y expr=\thisrowno{1}/1000, col sep=comma]{\datatableone};

        \addplot[blue, solid, thick, 
                mark=triangle, mark size={3.0}, every mark/.append style={solid, fill=gray}, mark repeat=3, opacity=0.9
                ] 
            table [x=x_data, y expr=\thisrowno{1}/1000, col sep=comma]{\datatableoneirsNHAAsicstat};
    
        \addplot[black, solid, 
                mark=square, mark size={2.5}, every mark/.append style={solid, fill=gray}, mark repeat=3, opacity=0.7
                ] 
            table [x=x_data, y expr=\thisrowno{1}/1000, col sep=comma]{\datatablefive};
            
        \addplot[black, solid,  
                mark=o, mark size={2.5}, every mark/.append style={solid, fill=gray}, mark repeat=3, opacity=0.7
                ] 
            table [x=x_data, y expr=\thisrowno{1}/1000, col sep=comma]{\datatableeleven};
        
        \addplot[black, solid, thick, 
                mark=star, mark size={3.0}, every mark/.append style={solid, fill=gray}, mark repeat=3, opacity=0.7
                ] 
            table [x=x_data, y expr=\thisrowno{1}/1000, col sep=comma]{\datatableweightedsumrandompsbf};

        \addplot[black, solid, thick, 
                mark=pentagon, mark size={3.0}, every mark/.append style={solid, fill=gray}, mark repeat=3, opacity=0.7
                ] 
            table [x=x_data, y expr=\thisrowno{1}/1000, col sep=comma]{\datatablewexpOU};

    \end{axis}
\end{tikzpicture}
}
\end{subfigure}%
\begin{subfigure}[b]{0.495\linewidth}
\centering
\resizebox{\linewidth}{!}{
\begin{tikzpicture}[thick,scale=0.99]
    \begin{axis}[
        width=7cm,
        height=5.5cm,
        xmin=0,
        xmax=99,
        ymin=5,
        ymax=9.5,
        grid=major,
        xlabel={No. of Episodes},
        ylabel={DL rate (bps/Hz)},
        xlabel style={at={(0.50,0.03)}},
        ylabel style={at={(0.1,0.50)}},
        label style={font=\large},
        legend pos=south east,
        legend cell align={left},
        legend style={at={(1.0,0)}, fill opacity=0.7, draw opacity=1.0, text opacity=1.0, font=\normalsize}
        ]
        \addplot[red, solid, thick, 
                mark=diamond, mark size={2.5}, every mark/.append style={solid, fill=gray}, mark repeat=3,  opacity=0.9
                ] 
            table [x=x_data, y expr=\thisrowno{2}/1000, col sep=comma]{\datatableone};
        \addlegendentry{MSF-DRL-LSSIC};

        \addplot[blue, solid, thick, 
                mark=triangle, mark size={3.0}, every mark/.append style={solid, fill=gray}, mark repeat=3, opacity=0.9
                ] 
            table [x=x_data, y expr=\thisrowno{2}/1000, col sep=comma]{\datatableoneirsNHAAsicstat};
        \addlegendentry{MSF-DRL-HSIC};
             
        \addplot[black, solid,
                mark=square, mark size={2.5}, every mark/.append style={solid, fill=gray}, mark repeat=3,  opacity=0.7
                ] 
            table [x=x_data, y expr=\thisrowno{2}/1000, col sep=comma]{\datatablefive};
        \addlegendentry{PerfCSI-DRL};
               
        \addplot[black, solid, 
                mark=o, mark size={2.5}, every mark/.append style={solid, fill=gray}, mark repeat=3,  opacity=0.7
                ] 
            table [x=x_data, y expr=\thisrowno{2}/1000, col sep=comma]{\datatableeleven};
        \addlegendentry{NoisCSI-DRL};
           
         \addplot[black, solid, thick, 
                mark=star, mark size={3.0}, every mark/.append style={solid, fill=gray}, mark repeat=3, opacity=0.7
                ] 
            table [x=x_data, y expr=\thisrowno{2}/1000, col sep=comma]{\datatableweightedsumrandompsbf};
        \addlegendentry{RandPSBF};

        \addplot[black, solid, thick, 
                mark=pentagon, mark size={3.0}, every mark/.append style={solid, fill=gray}, mark repeat=3, opacity=0.7
                ] 
            table [x=x_data, y expr=\thisrowno{2}/1000, col sep=comma]{\datatablewexpOU};
        \addlegendentry{OUPSBF};

    \end{axis}
\end{tikzpicture}
}
\end{subfigure}
\caption{Rate evolution in the static UE scenario.}
\vspace{-3mm}
\label{fig:rateevo_03ue_bs04x04_beamformerstat}
\end{figure}
\fi

\subsection{Study of MSF-DRL with LSSIC and HSIC}
The UL rate and DL rate evolution during the learning phase of the proposed algorithms MSF-DRL-(LSSIC/HSIC) and other benchmarks such as PerfCSI-DRL, NoisCSI-DRL, and OUPSBF are demonstrated in Fig. \ref{fig:rateevo_03ue_bs04x04_beamformerstat} for the ULA type BS with $M_t=10$ transmitting and $M_r=10$ receiving antennas (legends are shared across both the figures). These results are with static users. Both RIS1 and RIS2 have a total of $N_1=N_2=36$ reflecting elements ($N_{1h}=6$, $N_{1v}=6$, $N_{2h}=6$, $N_{2v}=6$). The performance of any intelligent agent is lower bounded by the performance of the \emph{RandPSBF} that randomly predicts RIS phases and beamformers. Initially, learning of both the OUPSBF and the proposed learning agents starts with a performance similar to that of the RandPSBF agent. This similarity is anticipated because in the beginning, DRL agents are initialized with random weights. Hence, their actions are also random.

The performance of the proposed methods is upper bounded by the performance of the semi-oracle, \emph{PerfCSI-DRL} method that uses complete knowledge of CSI, including the residual SI, to find the beamformers. The performance of PerfCSI-DRL starts with a very high value, and the convergence is faster, too. The initial higher data rates for the PerfCSI-DRL method can be attributed to using fixed beamformers, calculated using prior known CSI and residual SI. These beamformers are suitable for canceling residual SI, leading to higher data rates observed. In scenarios without the residual SI, the agent experiences only inter-user and inter-RIS interference. Despite beginning the training with fixed beamformers, we observe a continual increase in data rates for PerfCSI-DRL. This trend indicates that the agent is learning to mitigate inter-user and inter-RIS interference and enhance the signal component, thereby illustrating the impact of inter-user and inter-RIS interference in learning. In contrast to PerfCSI-DRL, the data rates for MSF-DRL-LSSIC begin at a very low value, suggesting that the initially predicted beamformers are poor (as the agent is not yet trained and predicts random actions). However, there has been a rapid increase in data rates since then, indicating excellent learning and faster convergence with LS-based SI cancellation. Eventually, the MSF-DRL-LSSIC method nearly matches the data rates achieved by the PerfCSI-DRL method.

A semi-oracle agent in a practical scenario is impossible; however, the aim is to reach as close to this as possible. Using MSF-DRL, one can avoid estimating CSI and reduce the overhead. For example, for $10$ BS antenna elements and $36$ RIS elements per RIS, the number of instantaneous CSI that must be input to the algorithm is $813$. In contrast, the proposed CSI oblivious MSF-DRL method requires to transmit only one pilot signal and receive only a scalar reward value. Note that when there is inter-RIS interference, CSI-based methods for canceling residual SI by beamformers will not perform well since they would require knowledge of two more channels, namely $\mathbf{g}_{IU}$ and $\mathbf{f}_{DI}$, which one would not estimate in practice. On the other hand, since we do not require CSI knowledge, we still perform well.

When the estimated CSI is noisy, the \emph{NoisCSI-DRL} method with the normalized mean square error (NMSE) between the actual and estimated CSI as $0$ dB \cite{wen2018deep} does not perform well. The estimate of residual SI channel $\mathbf{H}_{AA}^{}$ is also assumed to be noisy, with a noise variance of $10^{-12}$. As training progresses, we can observe that the performance of all the DRL methods increases. The rate evolution clearly shows the advantage of the proposed neural architecture and the action space exploration for both RIS phases and beamformers. Comparing OUPSBF with MSF-DRL-LSSIC and MSF-DRL-HSIC, we can claim that the proposed Gaussian action noise in MSF-DRL-(LSSIC/HSIC) helps to converge faster than the OU noise. However, with a higher value of the initial SD of the action noise, the learning becomes unstable, indicating the power of the right exploration strategy. With sufficient training, the MSF-DRL method approaches the performance of semi-oracle PerfCSI-DRL. When the estimated $\mathbf{H}_{AA}$ has a higher noise variance, then the MSF-DRL-HSIC agent cannot learn. Therefore, unless a good estimate of $\mathbf{H}_{AA}$ is available, the proposed MSF-DRL-LSSIC using a single pilot is better as it has the minimum signaling overhead.

\ifCLASSOPTIONtwocolumn
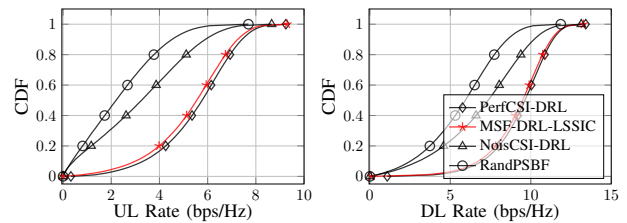
\begin{figure}[t]
\centering
    
\begin{subfigure}{.495\linewidth}
        \resizebox{\linewidth}{!}{
            \pgfplotstableread[col sep = comma]{datacdf/data_01UEUL_01UEDL_BS08_rateCdf_wsic_staticUE_wexpltnBF_LSSIC_urban_lowpathloss_BS10_RIS36_UL.csv}\datatablebuilUL

            \tikzstyle{mark_style} = [mark size={3.0}, mark repeat=20, mark phase=1]
             
              \pgfplotstableread[col sep = comma]{datacdf/data_01UEUL_01UEDL_BS08_rateCdf_wsic_staticUE_MRCZF_urban_lowpathloss_BS10_RIS36_UL.csv}\datatablebuilULMRCZF
            
              \pgfplotstableread[col sep = comma]{datacdf/data_01UEUL_01UEDL_BS08_rateCdf_wsic_staticUE_MRCZF_noisyCSIpercent100_urban_lowpathloss_BS10_RIS36_UL.csv}\datatablebuilULMRCZFnoisy

              \pgfplotstableread[col sep = comma]{datacdf/data_01UEUL_01UEDL_BS08_rateCdf_wsic_staticUE_randompsbf_B10R36_UL.csv}\datatablebuilULrandpsbf
               
\begin{tikzpicture}[thick,scale=0.99]
    \begin{axis}[
        width=7cm,
        height=5.5cm,
        xmin=-0.05,
        xmax=10,
        ymin=-0.05,
        ymax=+1.1,
        grid=major,
        xlabel={UL Rate (bps/Hz)},
        ylabel={CDF},
        xlabel style={at={(0.50,0.05)}},
        ylabel style={at={(0.06,0.50)}},
        ytick={0.0,0.2,...,1.0},
        label style={font=\large},
        legend pos=north west,
        legend cell align={left},
        legend style={fill opacity=0.6, draw opacity=1.0, text opacity=1.0, font=\normalsize}
        ]

        \addplot[black, solid, thick, mark=diamond, mark size={3.0}, mark_style, opacity=0.8
                ] 
            table [y=cdf, x=ddpg_df_2x128_gamma060_10000-100000, col sep=comma]{\datatablebuilULMRCZF};

        \addplot[red, solid, thick, 
                mark=star, mark size={3.0}, mark size={3.0}, mark_style, opacity=0.8
                ] 
            table [y=cdf, x=ddpg_df_2x128_gamma060_10000-100000, col sep=comma]{\datatablebuilUL};

        \addplot[black, solid, thick, mark=triangle, mark size={3.0}, mark_style, opacity=0.8
                ] 
            table [y=cdf, x=ddpg_df_2x128_gamma060_10000-100000, col sep=comma]{\datatablebuilULMRCZFnoisy};

        \addplot[black, solid, thick, mark=o, mark size={3.0}, mark_style, opacity=0.8
                ] 
            table [y=cdf, x=ddpg_df_2x128_gamma060_10000-100000, col sep=comma]{\datatablebuilULrandpsbf};

    \end{axis}
\end{tikzpicture}
        }
    \end{subfigure}\hspace{-3mm}%
     \begin{subfigure}{.495\linewidth}
        \resizebox{\linewidth}{!}{
            \pgfplotstableread[col sep = comma]{datacdf/data_01UEUL_01UEDL_BS08_rateCdf_wsic_staticUE_wexpltnBF_LSSIC_urban_lowpathloss_BS10_RIS36_DL.csv}\datatablebuilDL

            \tikzstyle{mark_style} = [mark size={3.0}, mark repeat=20, mark phase=1]
             
              \pgfplotstableread[col sep = comma]{datacdf/data_01UEUL_01UEDL_BS08_rateCdf_wsic_staticUE_MRCZF_urban_lowpathloss_BS10_RIS36_DL.csv}\datatablebuilDLMRCZF
            
              \pgfplotstableread[col sep = comma]{datacdf/data_01UEUL_01UEDL_BS08_rateCdf_wsic_staticUE_MRCZF_noisyCSIpercent100_urban_lowpathloss_BS10_RIS36_DL.csv}\datatablebuilDLMRCZFnoisy

              \pgfplotstableread[col sep = comma]{datacdf/data_01UEUL_01UEDL_BS08_rateCdf_wsic_staticUE_randompsbf_B10R36_DL.csv}\datatablebuilDLrandpsbf

\begin{tikzpicture}[thick,scale=0.99]
    \begin{axis}[
        width=7cm,
        height=5.5cm,
        xmin=-0.05,
        xmax=15,
        ymin=-0.05,
        ymax=+1.1,
        grid=major,
        xlabel={DL Rate (bps/Hz)},
        ylabel={CDF},
        xlabel style={at={(0.50,0.05)}},
        ylabel style={at={(0.06,0.50)}},
        ytick={0.0,0.2,...,1.0},
        label style={font=\large},
        legend pos=south east,
        legend cell align={left},
        legend style={fill opacity=0.6, draw opacity=1.0, text opacity=1.0, font=\normalsize}
        ]

        \addplot[black, solid, thick, mark=diamond, mark size={3.0}, mark_style, opacity=0.8
                ] 
            table [y=cdf, x=ddpg_df_2x128_gamma060_10000-100000, col sep=comma]{\datatablebuilDLMRCZF};
        \addlegendentry{PerfCSI-DRL};

        \addplot[red, solid, thick, 
                mark=star, mark size={3.0}, mark size={3.0}, mark_style, opacity=0.8
                ] 
            table [y=cdf, x=ddpg_df_2x128_gamma060_10000-100000, col sep=comma]{\datatablebuilDL};
        \addlegendentry{MSF-DRL-LSSIC};

        \addplot[black, solid, thick, mark=triangle, mark size={3.0}, mark_style, opacity=0.8
                ] 
            table [y=cdf, x=ddpg_df_2x128_gamma060_10000-100000, col sep=comma]{\datatablebuilDLMRCZFnoisy};
        \addlegendentry{NoisCSI-DRL};

        \addplot[black, solid, thick, mark=o, mark size={3.0}, mark_style, opacity=0.8
                ] 
            table [y=cdf, x=ddpg_df_2x128_gamma060_10000-100000, col sep=comma]{\datatablebuilDLrandpsbf};
        \addlegendentry{RandPSBF}; 
        
    \end{axis}
\end{tikzpicture}
        }
    \end{subfigure}%
\caption{CDF of observed rates.}
\label{fig:ratecdf_03ue}
\hspace{-3mm}
\end{figure}
\fi

\ifCLASSOPTIONtwocolumn
\begin{figure}[t]
    \centering
    \begin{subfigure}{.495\linewidth}
        \resizebox{\linewidth}{!}{
            \pgfplotstableread[col sep = comma]{datacdf/data_01UEUL_01UEDL_BS08_rateCdf_wsic_staticUE_wexpltnBF_LSSIC_urban_lowpathloss_BS6_RIS36_UL.csv}\datatablebuilULsix
            
            \pgfplotstableread[col sep = comma]{datacdf/data_01UEUL_01UEDL_BS08_rateCdf_wsic_staticUE_wexpltnBF_LSSIC_urban_lowpathloss_BS10_RIS36_UL.csv}\datatablebuilULbs

            \pgfplotstableread[col sep = comma]{datacdf/data_01UEUL_01UEDL_BS08_rateCdf_wsic_staticUE_wexpltnBF_LSSIC_urban_lowpathloss_BS14_RIS36_UL.csv}\datatablebuilULfourteen

            \tikzstyle{mark_style} = [mark size={3.0}, mark repeat=20, mark phase=1]            
               
\begin{tikzpicture}[thick,scale=0.99]
    \begin{axis}[
        width=7cm,
        height=5.5cm,
        xmin=-0.05,
        xmax=10,
        ymin=-0.05,
        ymax=+1.1,
        grid=major,
        xlabel={UL Rate (bps/Hz)},
        ylabel={CDF},
        xlabel style={at={(0.50,0.05)}},
        ylabel style={at={(0.05,0.50)}},
        label style={font=\large},
        ytick={0.0,0.2,...,1.0},
        legend pos=north west,
        legend cell align={left},
        legend style={fill opacity=0.6, draw opacity=1.0, text opacity=1.0, font=\tiny}
        ]
        
        \addplot[black, solid, thick, 
                mark=o, mark size={3.0}, mark size={3.0}, mark_style, opacity=0.8
                ] 
            table [y=cdf, x=ddpg_df_2x128_gamma060_10000-100000, col sep=comma]{\datatablebuilULsix};

        \addplot[red, solid, thick, 
                mark=star, mark size={3.0}, mark size={3.0}, mark_style, opacity=0.8
                ] 
            table [y=cdf, x=ddpg_df_2x128_gamma060_10000-100000, col sep=comma]{\datatablebuilULbs};

        \addplot[blue, solid, thick, 
                mark=diamond, mark size={3.0}, mark size={3.0}, mark_style, opacity=0.8
                ] 
            table [y=cdf, x=ddpg_df_2x128_gamma060_10000-100000, col sep=comma]{\datatablebuilULfourteen};

    \end{axis}
\end{tikzpicture}
        }
    \end{subfigure}\hspace{-3mm}%
    \begin{subfigure}{.495\linewidth}
        \resizebox{\linewidth}{!}{
            \pgfplotstableread[col sep = comma]{datacdf/data_01UEUL_01UEDL_BS08_rateCdf_wsic_staticUE_wexpltnBF_LSSIC_urban_lowpathloss_BS6_RIS36_DL.csv}\datatablebuilDLsix
            
            \pgfplotstableread[col sep = comma]{datacdf/data_01UEUL_01UEDL_BS08_rateCdf_wsic_staticUE_wexpltnBF_LSSIC_urban_lowpathloss_BS10_RIS36_DL.csv}\datatablebuilDLbs

            \pgfplotstableread[col sep = comma]{datacdf/data_01UEUL_01UEDL_BS08_rateCdf_wsic_staticUE_wexpltnBF_LSSIC_urban_lowpathloss_BS14_RIS36_DL.csv}\datatablebuilDLfourteen

            \tikzstyle{mark_style} = [mark size={3.0}, mark repeat=20, mark phase=1]

\begin{tikzpicture}[thick,scale=0.99]
    \begin{axis}[
        width=7cm,
        height=5.5cm,
        xmin=-0.05,
        xmax=15,
        ymin=-0.05,
        ymax=+1.1,
        grid=major,
        xlabel={DL Rate (bps/Hz)},
        ylabel={CDF},
        xlabel style={at={(0.50,0.05)}},
        ylabel style={at={(0.05,0.50)}},
        label style={font=\large},
        ytick={0.0,0.2,...,1.0},
        legend pos=south east,
        legend cell align={left},
        legend style={fill opacity=0.6, draw opacity=1.0, text opacity=1.0, font=\normalsize}
        ]
        
        \addplot[black, solid, thick, 
                mark=o, mark size={3.0}, mark size={3.0}, mark_style, opacity=0.8
                ] 
            table [y=cdf, x=ddpg_df_2x128_gamma060_10000-100000, col sep=comma]{\datatablebuilDLsix};
        \addlegendentry{$M_t=M_r=6$};

        \addplot[red, solid, thick, 
                mark=star, mark size={3.0}, mark size={3.0}, mark_style, opacity=0.8
                ] 
            table [y=cdf, x=ddpg_df_2x128_gamma060_10000-100000, col sep=comma]{\datatablebuilDLbs};
        \addlegendentry{$M_t=M_r=10$};

        \addplot[blue, solid, thick, 
                mark=diamond, mark size={3.0}, mark size={3.0}, mark_style, opacity=0.8
                ] 
            table [y=cdf, x=ddpg_df_2x128_gamma060_10000-100000, col sep=comma]{\datatablebuilDLfourteen};
        \addlegendentry{$M_t=M_r=14$};

    \end{axis}
\end{tikzpicture}
        }
    \end{subfigure}
    \caption{Effect of BS antenna elements.}
    \vspace{-3mm}
    \label{fig:ratecdf_03ue_BSelements}
\end{figure}
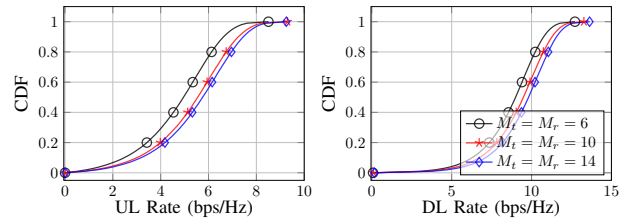
\fi

The cumulative distribution of UL and DL rates achieved by PerfCSI-DRL, NoisCSI-DRL, and MSF-DRL methods are given in Fig. \ref{fig:ratecdf_03ue}. We simulated $10000$ observations ($10$ episodes) for each method, and the data aggregated based on these observations are plotted. Note that the trained DRL agents are used to get the rate distribution. In Fig. \ref{fig:ratecdf_03ue_BSelements}, we observe that the UL and DL rates of MSF-DRL-LSSIC increase as the number of BS elements increases; this is because, as the number of antennas increases, the beam width can be narrowed.

\subsection{Performance with moving users}
To further illustrate the utility of our method, we even consider a scenario where the users are moving, and therefore, the LOS part of the Rician channels between RISs and UEs change over time. Usually, in practical scenarios, the CSI available to the agent is noisy, and the corresponding method is denoted as `NoisCSI-DRL". An NMSE of $-20$dB indicates that the noise in the estimated CSI has a power of $10\%$ of the power of that channel. The performance degrades when the NMSE of the NoisCSI-DRL method is higher, as shown in Fig. \ref{fig:rateevo_03ue_bs04x04_mobility_noisCSI}. As the proposed method MSF-DRL-LSSIC does not depend on CSI, its performance remains unaffected even if the CSI is noisy. Moreover, the proposed MSF-DRL-LSSIC performs very closely to the PerfCSI-DRL method. With the positional information about the UE as the observation from the environment \cite{evmorfos2022deep}, the MSF-DRL-LSSIC-pos performs slightly better than MSF-DRL-LSSIC. One may go for MSF-DRL-LSSIC-pos if the positional information is available.

\ifCLASSOPTIONtwocolumn
\begin{figure}[!t]
    \centering

    \pgfplotstableread[col sep = comma]{./datainterris/data_01UEUL_01UEDL_BS08_rateEvo_wsic_movingUE_wexpltnBF_LSSIC_urban_lowpathloss_speed1.0.csv}\datatableoneLSSICmoving

    \pgfplotstableread[col sep = comma]{./datainterris/data_01UEUL_01UEDL_BS08_rateEvo_wsic_movingUE_wexpltnBF_LSSIC_urban_lowpathloss_speed1.0_pos.csv}\datatableoneLSSICmovingpos

    \pgfplotstableread[col sep = comma]{./datainterris/data_01UEUL_01UEDL_BS08_rateEvo_wsic_movingUE_FixedBFMRCZF_urban_lowpathloss_speed1.0.csv}\datatableoneMRCMRTmoving

    \pgfplotstableread[col sep = comma]{./datainterris/data_01UEUL_01UEDL_BS08_rateEvo_wsic_movingUE_FixedBFMRCZF_urban_lowpathloss_speed1.0.csv}\datatableoneMRCMRTmoving
    
     \pgfplotstableread[col sep = comma]{./datainterris/data_01UEUL_01UEDL_BS08_rateEvo_wsic_movingUE_FixedBFMRCZF_noisyCSIpercent10_urban_lowpathloss_speed1.0.csv}\datatableoneMRCMRTmovingfive

     \pgfplotstableread[col sep = comma]{./datainterris/data_01UEUL_01UEDL_BS08_rateEvo_wsic_movingUE_FixedBFMRCZF_noisyCSIpercent100_urban_lowpathloss_speed1.0.csv}\datatableoneMRCMRTmovingone

        \begin{subfigure}[b]{0.495\linewidth}
        \centering
        \resizebox{\linewidth}{!}{
\begin{tikzpicture}[thick,scale=0.99]
    \begin{axis}[
        width=7cm,
        height=5.5cm,
        xmin=0,
        xmax=49,
        ymin=0,
        ymax=6,
        grid=major,
        xlabel={No. of Episodes},
        ylabel={UL rate (bps/Hz)},
        xlabel style={at={(0.50,0.0)}},
        ylabel style={at={(0.1,0.50)}},
        label style={font=\large},
        legend pos=south east,
        legend cell align={left},
        legend style={fill opacity=0.6, draw opacity=1.0, text opacity=1.0, font=\small}
        ]


        \addplot[black, solid, thick, 
                mark=triangle, mark size={3.0}, every mark/.append style={solid, fill=gray}, mark repeat=6, opacity=0.9
                ] 
            table [x=x_data, y expr=\thisrowno{1}/2000, col sep=comma]{\datatableoneLSSICmoving};

        \addplot[red, solid, thick, 
                mark=star, mark size={3.0}, every mark/.append style={solid, fill=gray}, mark repeat=6, opacity=0.7
                ] 
            table [x=x_data, y expr=\thisrowno{1}/2000, col sep=comma]{\datatableoneLSSICmovingpos};

        \addplot[blue, dashed, thick, 
                mark=o, mark size={3.0}, every mark/.append style={solid, fill=gray}, mark repeat=6, opacity=0.7
                ] 
            table [x=x_data, y expr=\thisrowno{1}/2000, col sep=comma]{\datatableoneMRCMRTmoving};
            
        \addplot[blue, solid, thick, 
            mark=square, mark size={3.0}, every mark/.append style={solid, fill=gray}, mark repeat=6, opacity=0.7
            ] 
        table [x=x_data, y expr=\thisrowno{1}/2000, col sep=comma]{\datatableoneMRCMRTmovingfive};

        \addplot[blue, solid, thick, 
            mark=diamond, mark size={3.0}, every mark/.append style={solid, fill=gray}, mark repeat=6, opacity=0.7
            ] 
        table [x=x_data, y expr=\thisrowno{1}/2000, col sep=comma]{\datatableoneMRCMRTmovingone};

    \end{axis}
\end{tikzpicture}
        }
       \end{subfigure}%
       \begin{subfigure}[b]{0.495\linewidth}
        \centering
        \resizebox{\linewidth}{!}{
\begin{tikzpicture}[thick,scale=0.99]
    \begin{axis}[
        width=7cm,
        height=5.5cm,
        xmin=0,
        xmax=49,
        ymin=0,
        ymax=10,
        grid=major,
        xlabel={No. of Episodes},
        ylabel={UL rate (bps/Hz)},
        xlabel style={at={(0.50,0.0)}},
        ylabel style={at={(0.1,0.50)}},
        label style={font=\large},
        legend pos=south east,
        legend cell align={left},
        legend style={fill opacity=0.6, draw opacity=1.0, text opacity=1.0, font=\normalsize}
        ]
            

        \addplot[black, solid, thick, 
                mark=triangle, mark size={3.0},every mark/.append style={solid, fill=gray}, mark repeat=6, opacity=0.9
                ] 
            table [x=x_data, y expr=\thisrowno{2}/2000, col sep=comma]{\datatableoneLSSICmoving};
        \addlegendentry{MSF-DRL-LSSIC};

        \addplot[red, solid, thick, 
                mark=star, mark size={3.0}, every mark/.append style={solid, fill=gray}, mark repeat=6, opacity=0.8
                ] 
            table [x=x_data, y expr=\thisrowno{2}/2000, col sep=comma]{\datatableoneLSSICmovingpos};
        \addlegendentry{MSF-DRL-LSSIC-pos};

        \addplot[blue, dashed, thick, 
                mark=o, mark size={3.0}, every mark/.append style={solid, fill=gray}, mark repeat=6, opacity=0.8
                ] 
            table [x=x_data, y expr=\thisrowno{2}/2000, col sep=comma]{\datatableoneMRCMRTmoving};
        \addlegendentry{PerfCSI-DRL};

        \addplot[blue, solid, thick, 
            mark=square, mark size={3.0}, every mark/.append style={solid, fill=gray}, mark repeat=6, opacity=0.8
            ] 
        table [x=x_data, y expr=\thisrowno{2}/2000, col sep=comma]{\datatableoneMRCMRTmovingfive};
        \addlegendentry{NoisCSI-DRL NMSE $-20$dB}

        \addplot[blue, solid, thick, 
            mark=diamond, mark size={3.0}, every mark/.append style={solid, fill=gray}, mark repeat=6, opacity=0.8
            ] 
        table [x=x_data, y expr=\thisrowno{2}/2000, col sep=comma]{\datatableoneMRCMRTmovingone};
        \addlegendentry{NoisCSI-DRL NMSE $0$dB}

    \end{axis}
\end{tikzpicture}
        }
       \end{subfigure}
        \caption{Rate evolution in the moving UE scenario.}
        \vspace{-3mm}
        \label{fig:rateevo_03ue_bs04x04_mobility_noisCSI}
\end{figure}
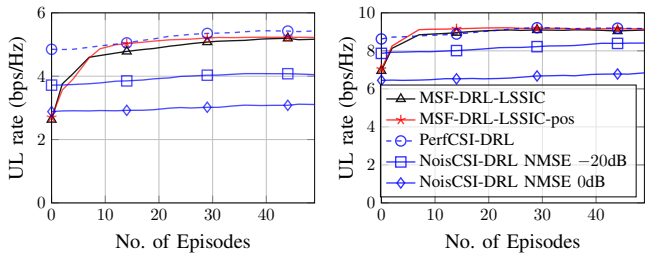
\fi

\ifCLASSOPTIONtwocolumn
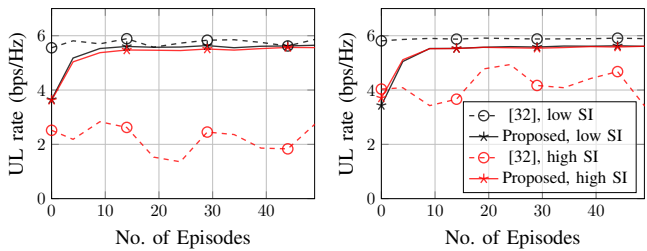
\begin{figure}[!t]
    \centering
    \pgfplotstableread[col sep = comma]{dataoneueoneris/data_01UEULDL_01UEDL_BS08_rateEvo_wsic_staticUE_Faisal_lowresidualSI.csv}\datatableoneMRCZFoneUEoneRISlowRSI

    \pgfplotstableread[col sep = comma]{dataoneueoneris/data_01UEULDL_01UEDL_BS08_rateEvo_wsic_staticUE_wexpltnBF_LSSIC_1UE1RIS_lowresidualSI.csv}\datatableoneLSSIConeUEoneRISlowRSI
    
    \pgfplotstableread[col sep = comma]{dataoneueoneris/data_01UEULDL_01UEDL_BS08_rateEvo_wsic_staticUE_Faisal_highresidualSI.csv}\datatableoneMRCZFoneUEoneRIShighRSI

    \pgfplotstableread[col sep = comma]{dataoneueoneris/data_01UEULDL_01UEDL_BS08_rateEvo_wsic_staticUE_wexpltnBF_LSSIC_1UE1RIS_highresidualSI.csv}\datatableoneLSSIConeUEoneRIShighRSI
           
        \begin{subfigure}[b]{0.495\linewidth}
        \centering
        \resizebox{\linewidth}{!}{
\begin{tikzpicture}[thick,scale=0.99]
    \begin{axis}[
        width=7cm,
        height=5.5cm,
        xmin=0,
        xmax=49,
        ymin=0,
        ymax=7,
        grid=major,
        xlabel={No. of Episodes},
        ylabel={UL rate (bps/Hz)},
        xlabel style={at={(0.50,0.0)}},
        ylabel style={at={(0.1,0.50)}},
        label style={font=\large},
        legend pos=south east,
        legend cell align={left},
        legend style={fill opacity=0.6, draw opacity=1.0, text opacity=1.0, font=\normalsize}
        ]

        \addplot[black, dashed, thick, 
                mark=o, mark size={3.0}, every mark/.append style={solid, fill=gray}, mark repeat=3, opacity=0.8
                ] 
            table [x=x_data, y expr=\thisrowno{1}/1000, col sep=comma]{\datatableoneMRCZFoneUEoneRISlowRSI};
        
         \addplot[black, solid, thick, 
                mark=star, mark size={3.0},every mark/.append style={solid, fill=gray}, mark repeat=3, opacity=0.8
                ] 
            table [x=x_data, y expr=\thisrowno{1}/1000, col sep=comma]{\datatableoneLSSIConeUEoneRISlowRSI};

        \addplot[red, dashed, thick, 
                mark=o, mark size={3.0}, every mark/.append style={solid, fill=gray}, mark repeat=3, opacity=0.8
                ] 
            table [x=x_data, y expr=\thisrowno{1}/1000, col sep=comma]{\datatableoneMRCZFoneUEoneRIShighRSI};
        
         \addplot[red, solid, thick, 
                mark=star, mark size={3.0},every mark/.append style={solid, fill=gray}, mark repeat=3, opacity=0.8
                ] 
            table [x=x_data, y expr=\thisrowno{1}/1000, col sep=comma]{\datatableoneLSSIConeUEoneRIShighRSI};

    \end{axis}
\end{tikzpicture}
        }
       \end{subfigure}%
       \begin{subfigure}[b]{0.495\linewidth}
        \centering
        \resizebox{\linewidth}{!}{
\begin{tikzpicture}[thick,scale=0.99]
    \begin{axis}[
        width=7cm,
        height=5.5cm,
        xmin=0,
        xmax=49,
        ymin=0,
        ymax=7,
        grid=major,
        xlabel={No. of Episodes},
        ylabel={UL rate (bps/Hz)},
        xlabel style={at={(0.50,0.0)}},
        ylabel style={at={(0.1,0.50)}},
        label style={font=\large},
        legend pos=south east,
        legend cell align={left},
        legend style={fill opacity=0.6, draw opacity=1.0, text opacity=1.0, font=\normalsize}
        ]

        \addplot[black, dashed, thick, 
                mark=o, mark size={3.0}, every mark/.append style={solid, fill=gray}, mark repeat=3, opacity=0.8
                ] 
            table [x=x_data, y expr=\thisrowno{2}/1000, col sep=comma]{\datatableoneMRCZFoneUEoneRISlowRSI};
        \addlegendentry{\cite{faisal2021deep}, low SI};
        
         \addplot[black, solid, thick, 
                mark=star, mark size={3.0},every mark/.append style={solid, fill=gray}, mark repeat=3, opacity=0.8
                ] 
            table [x=x_data, y expr=\thisrowno{2}/1000, col sep=comma]{\datatableoneLSSIConeUEoneRISlowRSI};
        \addlegendentry{Proposed, low SI};

        \addplot[red, dashed, thick, 
                mark=o, mark size={3.0}, every mark/.append style={solid, fill=gray}, mark repeat=3, opacity=0.8
                ] 
            table [x=x_data, y expr=\thisrowno{2}/1000, col sep=comma]{\datatableoneMRCZFoneUEoneRIShighRSI};
        \addlegendentry{\cite{faisal2021deep}, high SI};
        
         \addplot[red, solid, thick, 
                mark=star, mark size={3.0},every mark/.append style={solid, fill=gray}, mark repeat=3, opacity=0.8
                ] 
            table [x=x_data, y expr=\thisrowno{2}/1000, col sep=comma]{\datatableoneLSSIConeUEoneRIShighRSI};
        \addlegendentry{Proposed, high SI};

    \end{axis}
\end{tikzpicture}
        }
       \end{subfigure}
        \caption{MSF-DRL-LSSIC in system model with FD user.}
        \vspace{-3mm}
        \label{fig:rateevo_03ue_bs04x04_1UE1RIS_R1}
\end{figure}
\fi

\vspace{-3mm}

\subsection{Broad applicability of MSF-DRL-LSSIC}
The proposed two-stage DRL method is also applicable to other system models. For example, we consider a system model similar to \cite{faisal2021deep} in an urban environment with an FD BS serving one FD user with simultaneous UL and DL communication via one RIS. There is only one RIS, so the received signals do not have inter-RIS or inter-user interference. The authors of \cite{faisal2021deep} have used a semi-closed form solution to find the transmit beamformers, which depends on the knowledge of $\mathbf{H}_{AA}$. They solve iteratively for RIS phase shifts (by DRL approach) and transmit beamformers. Therefore, the quality of the solution of RIS phase shifts depends on the quality of the beamformers. The method suffers convergence issues when the residual SI is high. For example, the beamformers calculated using a semi-closed form solution given in \cite{faisal2021deep} cannot find good RIS phases in the presence of high residual SI ($\sigma_{AA}^2>10^{-6}$) even though the residual SI channel $\mathbf{H}_{AA}$ is assumed to be known. We have compared our MSF-DRL-LSSIC method with \cite{faisal2021deep} in Fig. \ref{fig:rateevo_03ue_bs04x04_1UE1RIS_R1}. The authors of \cite{faisal2021deep} showed results for a low residual SI of  $\sigma_{AA}^2=31.6*10^{-9}$, whereas we compare them also in the presence of high residual SI with $\sigma_{AA}^2=10^{-2}$. In low residual SI, MSF-DRL-LSSIC performs as well as \cite{faisal2021deep}. Note that \cite{faisal2021deep} still uses known CSI and $\mathbf{H}_{AA}$ to find the beamformers. However, in the presence of high residual SI, the performance of \cite{faisal2021deep} degrades, whereas the proposed method performs well. In the presence of high residual SI, the semi-closed form solution of FD beamforming does not help in the optimization method and does not converge as well as expected. Additionally, the proposed method applies to simpler scenarios of FD systems without RIS or MIMO configurations. In such scenarios, where MIMO and RIS are absent, $\mathbf{w}_R=\mathbf{w}_T=1$ and the paths via the RISs do not exist. Consequently, in the SINR expression, only the transmit powers can be optimized to maximize the weighted sum of UL and DL SINR. This aspect highlights the versatility of our contribution and the DRL approach.

\ifCLASSOPTIONtwocolumn
\begin{figure}[!t]
\centering
\pgfplotstableread[col sep = comma]{dataphaseshifts/data_01UEUL_01UEDL_BS08_rateEvo_wsic_staticUE_wexpltnBF_LSSIC_shadowedurban_highpathloss_anothertest.csv}\datapscont
\tikzstyle{mark_style} = []

\pgfplotstableread[col sep = comma]{dataphaseshifts/data_01UEUL_01UEDL_BS08_rateEvo_wsic_staticUE_wexpltnBF_LSSIC_shadowedurban_highpathloss_l2_g36_anothertest.csv}\datapstwo
\tikzstyle{mark_style} = []

\pgfplotstableread[col sep = comma]{dataphaseshifts/data_01UEUL_01UEDL_BS08_rateEvo_wsic_staticUE_wexpltnBF_LSSIC_shadowedurban_highpathloss_l4_g36_anothertest.csv}\datapsfour
\tikzstyle{mark_style} = []

\pgfplotstableread[col sep = comma]{dataphaseshifts/data_01UEUL_01UEDL_BS08_rateEvo_wsic_staticUE_wexpltnBF_LSSIC_shadowedurban_highpathloss_l8_g36_anothertest.csv}\datapssixteen
\tikzstyle{mark_style} = []

    \begin{subfigure}[b]{0.495\linewidth}
    \centering
    \resizebox{\linewidth}{!}{
    \begin{tikzpicture}[thick,scale=0.99]
    \begin{axis}[
        width=7cm,
        height=5.5cm,
        xmin=0,
        xmax=99,
        ymin=0,
        ymax=6.5,
        grid=major,
        xlabel={No. of Episodes},
        ylabel={UL rate (bps/Hz)},
        xlabel style={at={(0.50,0.05)}},
        ylabel style={at={(0.1,0.50)}},
        label style={font=\large},
        legend pos=south east,
        legend cell align={left},
        legend style={fill  opacity=0.7, draw opacity=1.0, text opacity=1.0, font=\footnotesize, anchor=south west}
        ]


        \addplot[black, solid, thick, 
                mark=square, mark size={2.5}, every mark/.append style={solid, fill=gray}, mark repeat=4, opacity=0.7
                ] 
            table [x=x_data, y expr=\thisrowno{1}/1000, col sep=comma]{\datapstwo};

        \addplot[black, solid, thick, 
                mark=star, mark size={3.0}, every mark/.append style={solid, fill=gray}, mark repeat=6, opacity=1.0
                ] 
            table [x=x_data, y expr=\thisrowno{1}/1000, col sep=comma]{\datapsfour};


        \addplot[black, solid, thick, 
                mark=pentagon, mark size={3.0}, every mark/.append style={solid, fill=gray}, mark repeat=5, opacity=0.7
                ] 
            table [x=x_data, y expr=\thisrowno{1}/1000, col sep=comma]{\datapssixteen};
            
        \addplot[black, dashed, thick, 
                mark=triangle, mark size={3.0}, every mark/.append style={solid, fill=gray}, mark repeat=3,  opacity=0.7
                ] 
            table [x=x_data, y expr=\thisrowno{1}/1000, col sep=comma]{\datapscont};

    \end{axis}
\end{tikzpicture}
    }
    \end{subfigure}%
    \begin{subfigure}[b]{0.495\linewidth}
    \centering
    \resizebox{\linewidth}{!}{
\begin{tikzpicture}[thick,scale=0.99]
    \begin{axis}[
        width=7cm,
        height=5.5cm,
        xmin=0,
        xmax=99,
        ymin=0,
        ymax=10,
        grid=major,
        xlabel={No. of Episodes},
        ylabel={DL rate (bps/Hz)},
        xlabel style={at={(0.50,0.05)}},
        ylabel style={at={(0.1,0.50)}},
        label style={font=\large},
        legend pos=south east,
        legend cell align={left},
        legend style={ fill opacity=0.7, draw opacity=1.0, text opacity=1.0, font=\normalsize}
        ]

        
        \addplot[black, solid, thick, 
                mark=square, mark size={2.5}, every mark/.append style={solid, fill=gray}, mark repeat=4, opacity=0.7
                ] 
            table [x=x_data, y expr=\thisrowno{2}/1000, col sep=comma]{\datapstwo};
        \addlegendentry{n=1};
        
        \addplot[black, solid, thick, 
                mark=star, mark size={3.0}, every mark/.append style={solid, fill=gray}, mark repeat=6, opacity=1.0
                ] 
            table [x=x_data, y expr=\thisrowno{2}/1000, col sep=comma]{\datapsfour};
        \addlegendentry{n=2};
        

        \addplot[black, solid, thick, 
                mark=pentagon, mark size={3.0}, every mark/.append style={solid, fill=gray}, mark repeat=5, opacity=0.7
                ] 
            table [x=x_data, y expr=\thisrowno{2}/1000, col sep=comma]{\datapssixteen};
        \addlegendentry{n=3};

         \addplot[black, dashed, thick, 
                mark=triangle, mark size={3.0}, every mark/.append style={solid, fill=gray}, mark repeat=3, opacity=0.7
                ] 
            table [x=x_data, y expr=\thisrowno{2}/1000, col sep=comma]{\datapscont};
        \addlegendentry{Cont};
    \end{axis}
\end{tikzpicture}
    }
    \end{subfigure}
\caption{Effect of quantization on phase shifts of RISs on MSF(Q)-DRL methods. When $n=2$, the phase shifts can take $Q=2^n=4$ values $\{0, \pi/2, \pi, 3\pi/2\}$.}
\vspace{-3mm}
\label{fig:rateevo_03ue_bs04x04_PS}
\end{figure}
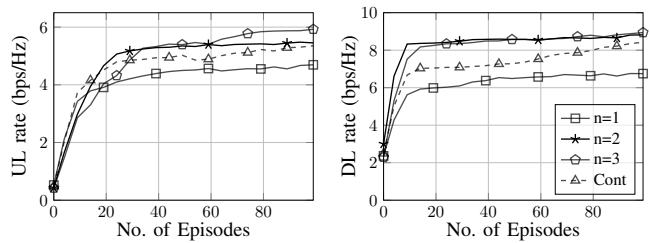 
\fi

\subsection{Leveraging DRL when only Limited Phase Information can be communicated}
\textbf{Quantized MSF-DRL:} We first experimented with an urban scenario where MSF(Q)-DRL performs as well as MSF-DRL. This experiment implies that with quantized phases, one can reduce the signaling \footnote{We have not included these results for space constraint.}. However, we observed even more interesting results when users are considered to be in a shadowed urban environment with $\alpha_{AU}=\alpha_R=4.5$ and slightly higher transmit powers with $p_U^{max}=0.2$ W and $p_A^{max}=3.16$ W for both MSF-DRL and MSF(Q)-DRL. In Fig. \ref{fig:rateevo_03ue_bs04x04_PS}, we observe that $n=2$ and $n=3$ bit quantization for the RIS phase shifts lead to reduced signaling as well as improved performance of the UL and DL data rates towards higher values compared to using continuous-valued phase shifts in the MSF-DRL method. Usually, the performance of any quantized system is worse than that of the continuous system. However, in a DRL framework with quantized phases, the performance is better than that of the continuous phase because the beamformers are adjusted so that the UL and DL rates are high even with the quantized phase angles. Furthermore, with quantized angles, the action space for RIS phases is quite small, resulting in faster convergence to better rates. At episode $15$, the MSF(Q)-DRL with $n=2$ starts outperforming the MSF-DRL. At episode $40$, the UL and DL rates of MSF(Q)-DRL for $n=2$ are better by $7.1\%$ and $22.28\%$, respectively, than the UL and DL rates for MSF-DRL method. Using $n=3$ bit quantization for the RIS phase shift values leads to better UL data rates than $2$-bit quantization. However, the convergence is slower for $n=3$ as the possible phase shift values, and hence, the action space increases with the increased number of bits. Even though the convergence of DL rates is faster for MSF(Q)-DRL for $n=2$ and $n=3$, they converge to almost the same values over time. With more training, the data rates of the continuous phase MSF-DRL method gradually improve but cannot outperform the MSF(Q)-DRL methods. It shows that the continuous phase values are unnecessary to learn if the beamformers are also learned jointly. The number of bits that need to be transmitted is $144$ for $2$-bit quantization ($Q=2^n=4$) compared to $4608$ bits for the MSF-DRL or the semi-oracle methods, as shown in Table. \ref{tab:runtime}. Therefore, the joint learning of RIS phases and beamformers using the DRL method helps significantly reduce the signaling between the BS and RIS. 

\ifCLASSOPTIONtwocolumn
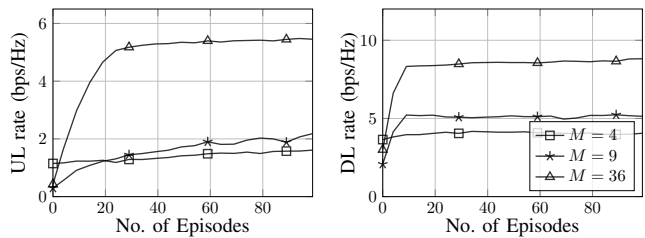
\begin{figure}[!t]
\centering
\pgfplotstableread[col sep = comma]{dataphaseshifts/data_01UEUL_01UEDL_BS08_rateEvo_wsic_staticUE_wexpltnBF_LSSIC_shadowedurban_highpathloss_l4_g4_anothertest.csv}\datagroupfour
\tikzstyle{mark_style} = []

\pgfplotstableread[col sep = comma]{dataphaseshifts/data_01UEUL_01UEDL_BS08_rateEvo_wsic_staticUE_wexpltnBF_LSSIC_shadowedurban_highpathloss_l4_g9_anothertest.csv}\datagroupnine
\tikzstyle{mark_style} = []

\pgfplotstableread[col sep = comma]{dataphaseshifts/data_01UEUL_01UEDL_BS08_rateEvo_wsic_staticUE_wexpltnBF_LSSIC_shadowedurban_highpathloss_l4_g36_anothertest.csv}\datapseight
\tikzstyle{mark_style} = []

\begin{subfigure}[b]{0.495\linewidth}
\centering
\resizebox{\linewidth}{!}{
\begin{tikzpicture}[thick,scale=0.8]
    \begin{axis}[
        width=7cm,
        height=5.5cm,
        xmin=0,
        xmax=99,
        ymin=0,
        ymax=6.5,
        grid=major,
        xlabel={No. of Episodes},
        ylabel={UL rate (bps/Hz)},
        xlabel style={at={(0.50,0.05)}},
        ylabel style={at={(0.1,0.50)}},
        label style={font=\large},
        legend pos=south east,
        legend cell align={left},
        legend style={fill opacity=0.8, draw opacity=1.0, text opacity=1.0, font=\normalsize}
        ]

        \addplot[black, solid, thick, 
                mark=square, mark size={2.5}, every mark/.append style={solid, fill=gray}, mark repeat=6, opacity=0.8
                ] 
            table [x=x_data, y expr=\thisrowno{1}/1000, col sep=comma]{\datagroupfour};

        \addplot[black, solid, thick, 
                mark=star, mark size={3.0}, every mark/.append style={solid, fill=gray}, mark repeat=6, opacity=0.8
                ] 
            table [x=x_data, y expr=\thisrowno{1}/1000, col sep=comma]{\datagroupnine};

        \addplot[black, solid, thick, 
                mark=triangle, mark size={3.0}, every mark/.append style={solid, fill=gray}, mark repeat=6, opacity=0.8
                ] 
            table [x=x_data, y expr=\thisrowno{1}/1000, col sep=comma]{\datapseight};

    \end{axis}
\end{tikzpicture}
}
\end{subfigure}%
\begin{subfigure}[b]{0.495\linewidth}
\centering
\resizebox{\linewidth}{!}{
\begin{tikzpicture}[thick,scale=0.8]
    \begin{axis}[
        width=7cm,
        height=5.5cm,
        xmin=0,
        xmax=99,
        ymin=0,
        ymax=12,
        grid=major,
        xlabel={No. of Episodes},
        ylabel={DL rate (bps/Hz)},
        xlabel style={at={(0.50,0.05)}},
        ylabel style={at={(0.1,0.50)}},
        label style={font=\large},
        legend pos=south east,
        legend cell align={left},
        legend style={ fill opacity=0.8, draw opacity=1.0, text opacity=1.0, font=\normalsize}
        ]
        
        \addplot[black, solid, thick, 
                mark=square, mark size={2.5}, every mark/.append style={solid, fill=gray}, mark repeat=6, opacity=0.8
                ] 
            table [x=x_data, y expr=\thisrowno{2}/1000, col sep=comma]{\datagroupfour};
        \addlegendentry{$M=4$};

        \addplot[black, solid, thick, 
                mark=star, mark size={3.0}, every mark/.append style={solid, fill=gray}, mark repeat=6, opacity=0.8
                ] 
            table [x=x_data, y expr=\thisrowno{2}/1000, col sep=comma]{\datagroupnine};
        \addlegendentry{$M=9$};

        \addplot[black, solid, thick, 
                mark=triangle, mark size={3.0}, every mark/.append style={solid, fill=gray}, mark repeat=6, opacity=0.8
                ] 
            table [x=x_data, y expr=\thisrowno{2}/1000, col sep=comma]{\datapseight};
        \addlegendentry{$M=36$};
        
    \end{axis}
\end{tikzpicture}
}
\end{subfigure}
\hspace{-3mm}
\caption{Effect of grouping passive elements of RISs}
\label{fig:rateevo_03ue_bs04x04_group}
\hspace{-3mm}
\end{figure} 
\fi

\textbf{Grouped Quantized MSF-DRL:} To reduce the signaling further,  GP(M)-MSF(Q)-DRL ($M=9$) groups learns only nine RIS phase shift values. If $M=36$, then all $36$ RIS phases will be distinct. It can be seen from Fig. \ref{fig:rateevo_03ue_bs04x04_group} that when all the RIS phases are learned, the performance is best, whereas when a lesser number of phases are learned, the performance degrades. However, using the proposed GP(M)-MSF(Q)-DRL method, it is possible to learn only $M$ phase shifts instead of $N$, which is otherwise difficult to solve with traditional optimization methods without changing the solution method completely. Efficiently grouping these RIS phase shifts is also an interesting problem to investigate and can be a part of future research. In the GP(M)-MSF(Q)-DRL method, with $2$-bit quantization and $M=9$ groups, the BS needs to transmit only $36$ bits to both the RISs, which is $128$ times less than the MSF-DRL method as shown in Table. \ref{tab:runtime}.

\ifCLASSOPTIONtwocolumn
\begin{table}[!t]
\caption{Proposed methods reduces signaling}
\label{tab:runtime}
  \begin{center}
    \begin{tabular}{|p{1.7cm}|p{1.3cm}|p{1.6cm}| p{2.6cm}|}
    \hline
    Methods & FLOPs& Needs CSI, SI ($\#$values) & Signaling from BS to RIS (in bits) \\
    \hline
     PerfCSI-DRL & $2.9\times 10^4$  & Yes ($813$) & $64N_1+64N_2=4608$\\
    \hline 
    MSF-DRL & $3.3\times 10^4$  & \textbf{No} ($2$) & $64N_1+64N_2=4608$ \\
    \hline 
   
    MSF(Q)-DRL &$5.46\times 10^4$ & \textbf{No} ($2$) & $2N_1+2N_2=\mathbf{144}$\\
    \hline 
    GP(M)-MSF(Q)-DRL & $5.46\times 10^4$ & \textbf{No} ($2$) & $2M+2M=\mathbf{36}$\\
    \hline 
\end{tabular}
\vspace{-3mm}
\end{center}
\end{table}
\fi

\subsection{Computational Analysis} 
Considering $2+N_1+N_2+4M_t+2$ as the dimension of the input to the FE, and there are $D$ neurons at every layer of the $2$-layer feature extractor network, the total number of floating point operations (FLOPs) at the feature extractor is $(2+N_1+N_2+4M_t+2)D+D^2$. The FLOPS for $SN_1$ and $SN_2$ for the RIS phases are $D\, N_1$ and $D\, N_2$, respectively. Regarding beamformers, the FLOPS of $SN_3$ and $SN_4$ are $2M_tD$; the same for $SN_5$ and $SN_6$ are $2M_rD$. The FLOPs for $SN_7$ and $SN_8$ for transmit powers are $2D$. The number of FLOPs for $SN_1$ and $SN_2$ change for the MSF(Q)-DRL; i.e. for a $n=2$, the total number of FLOPs for $SN_1$ and $SN_2$ are $D\, N_1\,2^n$ and $D\, N_2\,2^n$, respectively. PerfCSI-DRL does not predict beamformers using the DRL framework, so it has a slightly lesser FLOP count. However, PerfCSI-DRL requires a separate computation to process the CSI and find the beamformers. As the same architecture is used for both the MSF(Q)-DRL and GP(M)-MSF(Q)-DRL methods, they have the same complexity/FLOP count. The number of FLOPs for critic networks of both MSF-DRL and PerfCSI-DRL is $D(N_1+N_2+4M_t+2)+D^2+D$; the same for MSF(Q)-DRL and GP(M)-MSF(Q)-DRL methods is $D(N_1\,2^n+N_2\,2^n+4M_t+2)+D^2+D$. The complexity of the proposed methods is in the same order as PerfCSI-DRL, as shown in Table \ref{tab:runtime}.

    \section{Conclusion}
In this work, we propose MSF-DRL, a two-stage online DRL algorithm that maximizes the weighted sum rate to improve the performance of the UL and DL users. The proposed method can predict RIS phases, beamformers, and transmit powers in a full-duplex scenario without CSI knowledge or the exact knowledge of residual SI, i.e., the knowledge of $\mathbf{H}_{AA}$ matrix. Instead, it uses only one pilot signal at every time step to obtain an estimate of $\mathbf{H}_{AA}$ in the first stage, which cancels a part of the residual SI and instigates the learning. 
In the second stage, the part of SI not canceled in the first stage is attempted to cancel by the beamformers predicted by the DRL agent. The performance of the proposed method is very close to that of the semi-oracle DRL method that uses CSI. The proposed MSF(Q)-DRL method with $2$-bit quantization achieves better and faster convergence due to the reduced action space yet uses $32$ times lesser signaling than the MSF-DRL method. When the allowed signaling is reduced further, the RIS phases can be grouped and optimized with GP(M)-MSF(Q)-DRL, which needs $128$ times less signaling than the MSF-DRL method. 

To the best of our knowledge, this is the first work that predicts beamformers along with RIS phase shifts and transmit powers without any knowledge of CSI, and the method can handle significant residual SI. A possible future direction of practical interest will be the multi-user scenario. To serve multiple users simultaneously, one typically needs to perform spatial multiplexing \cite{sun2014mimo,jeon2017new}. Extending our work to learn spatial multiplexing from available observations is an interesting avenue for further research. One promising option can be to collect the SINRs from multiple users spread in a 2D space with the assistance of multiple BSs. A detailed investigation is out of the scope of this current work. However, this avenue has potential for further exploration and could significantly advance the field.

    \bibliographystyle{IEEEtran}
    \bibliography{library.bib}

\begin{thebibliography}{10}
\providecommand{\url}[1]{#1}
\csname url@samestyle\endcsname
\providecommand{\newblock}{\relax}
\providecommand{\bibinfo}[2]{#2}
\providecommand{\BIBentrySTDinterwordspacing}{\spaceskip=0pt\relax}
\providecommand{\BIBentryALTinterwordstretchfactor}{4}
\providecommand{\BIBentryALTinterwordspacing}{\spaceskip=\fontdimen2\font plus
\BIBentryALTinterwordstretchfactor\fontdimen3\font minus
  \fontdimen4\font\relax}
\providecommand{\BIBforeignlanguage}[2]{{%
\expandafter\ifx\csname l@#1\endcsname\relax
\typeout{** WARNING: IEEEtran.bst: No hyphenation pattern has been}%
\typeout{** loaded for the language `#1'. Using the pattern for}%
\typeout{** the default language instead.}%
\else
\language=\csname l@#1\endcsname
\fi
#2}}
\providecommand{\BIBdecl}{\relax}
\BIBdecl

\bibitem{di2020smart}
M.~Di~Renzo, A.~Zappone, M.~Debbah, M.-S. Alouini, C.~Yuen, J.~De~Rosny, and
  S.~Tretyakov, ``Smart radio environments empowered by reconfigurable
  intelligent surfaces: How it works, state of research, and the road ahead,''
  \emph{IEEE J. Sel. Areas Commun.}, vol.~38, no.~11, pp. 2450--2525, 2020.

\bibitem{abeywickrama2020intelligent}
S.~Abeywickrama, R.~Zhang, Q.~Wu, and C.~Yuen, ``Intelligent reflecting
  surface: Practical phase shift model and beamforming optimization,''
  \emph{IEEE Trans. Commun.}, vol.~68, no.~9, pp. 5849--5863, 2020.

\bibitem{wang2020jointpelian}
P.~Wang, J.~Fang, L.~Dai, and H.~Li, ``Joint transceiver and large intelligent
  surface design for massive mimo mmwave systems,'' \emph{IEEE Trans. Wireless
  Commun.}, vol.~20, no.~2, pp. 1052--1064, 2020.

\bibitem{gopi2020intelligent}
S.~Gopi, S.~Kalyani, and L.~Hanzo, ``Intelligent reflecting surface assisted
  beam index-modulation for millimeter wave communication,'' \emph{IEEE Trans.
  Wireless Commun.}, vol.~20, no.~2, pp. 983--996, 2020.

\bibitem{cheng2021downlink}
Y.~Cheng, K.~H. Li, Y.~Liu, K.~C. Teh, and H.~V. Poor, ``Downlink and uplink
  intelligent reflecting surface aided networks: {NOMA} and {OMA},'' \emph{IEEE
  Trans. Wireless Commun.}, vol.~20, no.~6, pp. 3988--4000, 2021.

\bibitem{li2020reconfigurable}
S.~Li, B.~Duo, X.~Yuan, Y.-C. Liang, and M.~Di~Renzo, ``Reconfigurable
  intelligent surface assisted uav communication: Joint trajectory design and
  passive beamforming,'' \emph{IEEE Wireless Commun. Lett.}, vol.~9, no.~5, pp.
  716--720, 2020.

\bibitem{charishma2021outage}
M.~Charishma, A.~Subhash, S.~Shekhar, and S.~Kalyani, ``Outage probability
  expressions for an {IRS}-assisted system with and without source-destination
  link for the case of quantized phase shifts in $\kappa$--$\mu$ fading,''
  \emph{IEEE Trans. Commun.}, vol.~70, no.~1, pp. 101--117, 2021.

\bibitem{kammoun2020asymptotic}
A.~Kammoun, A.~Chaaban, M.~Debbah, M.-S. Alouini \emph{et~al.}, ``Asymptotic
  max-min sinr analysis of reconfigurable intelligent surface assisted miso
  systems,'' \emph{IEEE Trans. Wireless Commun.}, vol.~19, no.~12, pp.
  7748--7764, 2020.

\bibitem{jayalal2022sinr}
L.~Jayalal, S.~Shekhar, A.~Subhash, and S.~Kalyani, ``{SINR} analysis of an
  {IRS} assisted {MU}-{MISO} system,'' \emph{arXiv preprint arXiv:2208.03664},
  2022.

\bibitem{du2021capacity}
L.~Du, J.~Ma, Q.~Liang, C.~Li, and Y.~Tang, ``Capacity characterization for
  reconfigurable intelligent surfaces assisted wireless communications with
  interferer,'' \emph{IEEE Trans. Commun.}, vol.~70, no.~3, pp. 1546--1558,
  2021.

\bibitem{huang2019reconfigurable}
C.~Huang, A.~Zappone, G.~C. Alexandropoulos, M.~Debbah, and C.~Yuen,
  ``Reconfigurable intelligent surfaces for energy efficiency in wireless
  communication,'' \emph{IEEE Trans. Wireless Commun.}, vol.~18, no.~8, pp.
  4157--4170, 2019.

\bibitem{wu2019intelligent}
Q.~Wu and R.~Zhang, ``Intelligent reflecting surface enhanced wireless network
  via joint active and passive beamforming,'' \emph{IEEE Trans. Wireless
  Commun.}, vol.~18, no.~11, pp. 5394--5409, 2019.

\bibitem{yu2019miso}
X.~Yu, D.~Xu, and R.~Schober, ``{MISO} wireless communication systems via
  intelligent reflecting surfaces,'' in \emph{Proc. IEEE/CIC International
  Conference on Communications in China (ICCC 2019)}, 2019, pp. 735--740.

\bibitem{jensen2020optimal}
T.~L. Jensen and E.~De~Carvalho, ``An optimal channel estimation scheme for
  intelligent reflecting surfaces based on a minimum variance unbiased
  estimator,'' in \emph{Proc. IEEE International Conference on Acoustics,
  Speech and Signal Processing (ICASSP 2020)}, May 2020, pp. 5000--5004.

\bibitem{sabharwal2014}
A.~Sabharwal, P.~Schniter, D.~Guo, D.~W. Bliss, S.~Rangarajan, and R.~Wichman,
  ``In-band full-duplex wireless: {C}hallenges and opportunities,'' \emph{IEEE
  J. Sel. Areas Commun.}, vol.~32, no.~9, pp. 1637--1652, 2014.

\bibitem{pan2021full}
G.~Pan, J.~Ye, J.~An, and M.-S. Alouini, ``Full-duplex enabled intelligent
  reflecting surface systems: Opportunities and challenges,'' \emph{IEEE
  Wireless Communications}, vol.~28, no.~3, pp. 122--129, 2021.

\bibitem{bharadia2013full}
D.~Bharadia, E.~McMilin, and S.~Katti, ``Full duplex radios,'' in
  \emph{Proceedings of the ACM SIGCOMM 2013 conference on SIGCOMM}, 2013, pp.
  375--386.

\bibitem{riihonen2011}
T.~Riihonen, S.~Werner, and R.~Wichman, ``Mitigation of loopback
  self-interference in full-duplex {MIMO} relays,'' \emph{IEEE Trans. Signal
  Process.}, vol.~59, no.~12, pp. 5983--5993, 2011.

\bibitem{everett2014passive}
E.~Everett, A.~Sahai, and A.~Sabharwal, ``Passive self-interference suppression
  for full-duplex infrastructure nodes,'' \emph{IEEE Trans. Wireless Commun.},
  vol.~13, no.~2, pp. 680--694, 2014.

\bibitem{smida2023full}
B.~Smida, A.~Sabharwal, G.~Fodor, G.~C. Alexandropoulos, H.~A. Suraweera, and
  C.-B. Chae, ``Full-duplex wireless for 6g: Progress brings new opportunities
  and challenges,'' \emph{IEEE J.Sel. Areas Commun.}, vol.~41, no.~9, pp.
  2729--2750, 2023.

\bibitem{cai2021intelligent}
Y.~Cai, M.-M. Zhao, K.~Xu, and R.~Zhang, ``Intelligent reflecting surface aided
  full-duplex communication: Passive beamforming and deployment design,''
  \emph{IEEE Trans. Wireless Commun.}, vol.~21, no.~1, pp. 383--397, 2022.

\bibitem{yang2021optimal}
Z.~Yang, C.~Huang, J.~Shi, Y.~Chau, W.~Xu, Z.~Zhang, and M.~Shikh-Bahaei,
  ``Optimal control for full-duplex communications with reconfigurable
  intelligent surface,'' in \emph{Proc. IEEE Intl. Conf. Commun. (ICC 2021)},
  2021, pp. 1--6.

\bibitem{zhang2020sum}
Y.~Zhang, C.~Zhong, Z.~Zhang, and W.~Lu, ``Sum rate optimization for two way
  communications with intelligent reflecting surface,'' \emph{IEEE Commun.
  Lett.}, vol.~24, no.~5, pp. 1090--1094, 2020.

\bibitem{nguyen2021cooperative}
B.~C. Nguyen, T.~M. Hoang, P.~T. Tran, T.~N. Nguyen, V.-D. Phan, B.~V. Minh,
  and M.~Voznak, ``Cooperative communications for improving the performance of
  bidirectional full-duplex system with multiple reconfigurable intelligent
  surfaces,'' \emph{IEEE Access}, vol.~9, pp. 134\,733--134\,742, 2021.

\bibitem{elhattab2021reconfigurable}
M.~Elhattab, M.~A. Arfaoui, C.~Assi, and A.~Ghrayeb, ``Reconfigurable
  intelligent surface enabled full-duplex/half-duplex cooperative
  non-orthogonal multiple access,'' \emph{IEEE Trans. Wireless Commun.},
  vol.~21, no.~5, pp. 3349--3364, 2021.

\bibitem{perera2022sum}
P.~P. Perera, V.~G. Warnasooriya, D.~Kudathanthirige, and H.~A. Suraweera,
  ``Sum rate maximization in {STAR}-{RIS} assisted full-duplex communication
  systems,'' in \emph{Proc. IEEE International Conference on Communications
  (ICC 2022)}, 2022, pp. 3281--3286.

\bibitem{peng2021multiuser}
Z.~Peng, Z.~Zhang, C.~Pan, L.~Li, and A.~L. Swindlehurst, ``Multiuser
  full-duplex two-way communications via intelligent reflecting surface,''
  \emph{IEEE Trans. Signal Process.}, vol.~69, pp. 837--851, 2021.

\bibitem{zhang2015full}
Z.~Zhang, X.~Chai, K.~Long, A.~V. Vasilakos, and L.~Hanzo, ``Full duplex
  techniques for 5{G} networks: self-interference cancellation, protocol
  design, and relay selection,'' \emph{IEEE Commun. Mag.}, vol.~53, no.~5, pp.
  128--137, 2015.

\bibitem{taha2020deep}
A.~Taha, Y.~Zhang, F.~B. Mismar, and A.~Alkhateeb, ``Deep reinforcement
  learning for intelligent reflecting surfaces: Towards standalone operation,''
  in \emph{Proc. IEEE 21st international workshop on signal processing advances
  in wireless communications (SPAWC 2020)}, 2020, pp. 1--5.

\bibitem{feng2020deep}
K.~Feng, Q.~Wang, X.~Li, and C.-K. Wen, ``Deep reinforcement learning based
  intelligent reflecting surface optimization for {MISO} communication
  systems,'' \emph{IEEE Wireless Commun. Lett.}, vol.~9, no.~5, pp. 745--749,
  2020.

\bibitem{lin2020deep}
J.~Lin, Y.~Zout, X.~Dong, S.~Gong, D.~T. Hoang, and D.~Niyato, ``Deep
  reinforcement learning for robust beamforming in {IRS}-assisted wireless
  communications,'' in \emph{Proc. IEEE Global Communications Conference
  ({GLOBECOM} 2020)}, 2020, pp. 1--6.

\bibitem{faisal2021deep}
A.~Faisal, I.~Al-Nahhal, O.~A. Dobre, and T.~M. Ngatched, ``Deep reinforcement
  learning for optimizing {RIS}-assisted {HD}-{FD} wireless systems,''
  \emph{IEEE Commun. Lett.}, vol.~25, no.~12, pp. 3893--3897, 2021.

\bibitem{subhash2022max}
A.~Subhash, A.~Kammoun, A.~Elzanaty, S.~Kalyani, Y.~H. Al-Badarneh, and M.-S.
  Alouini, ``Max-min sinr optimization for ris-aided uplink communications with
  green constraints,'' \emph{IEEE Wireless Commun. Lett.}, 2023.

\bibitem{shekhar2022instantaneous}
S.~Shekhar, A.~Subhash, T.~Kella, and S.~Kalyani, ``Instantaneous channel
  oblivious phase shift design for an {IRS}-assisted {SIMO} system with
  quantized phase shift,'' \emph{arXiv preprint arXiv:2211.03317}, 2022.

\bibitem{alwazani2020intelligent}
H.~Alwazani, A.~Kammoun, A.~Chaaban, M.~Debbah, M.-S. Alouini \emph{et~al.},
  ``Intelligent reflecting surface-assisted multi-user {MISO} communication:
  Channel estimation and beamforming design,'' \emph{IEEE Open J. Commun.
  Soc.}, vol.~1, pp. 661--680, 2020.

\bibitem{luo2021spatial}
S.~Luo, P.~Yang, Y.~Che, K.~Yang, K.~Wu, K.~C. Teh, and S.~Li, ``Spatial
  modulation for {RIS}-assisted uplink communication: Joint power allocation
  and passive beamforming design,'' \emph{IEEE Trans. Commun.}, vol.~69,
  no.~10, pp. 7017--7031, 2021.

\bibitem{mohammadi2015full}
M.~Mohammadi, H.~A. Suraweera, Y.~Cao, I.~Krikidis, and C.~Tellambura,
  ``Full-duplex radio for uplink/downlink wireless access with spatially random
  nodes,'' \emph{IEEE Trans. Commun.}, vol.~63, no.~12, pp. 5250--5266, 2015.

\bibitem{zhi2021statistical}
K.~Zhi, C.~Pan, H.~Ren, and K.~Wang, ``Statistical csi-based design for
  reconfigurable intelligent surface-aided massive mimo systems with direct
  links,'' \emph{IEEE Wireless Commun. Lett.}, vol.~10, no.~5, pp. 1128--1132,
  2021.

\bibitem{chen2021robust}
Y.~Chen, Y.~Wang, and L.~Jiao, ``Robust transmission for reconfigurable
  intelligent surface aided millimeter wave vehicular communications with
  statistical csi,'' \emph{IEEE Trans. Wireless Commun.}, vol.~21, no.~2, pp.
  928--944, 2021.

\bibitem{subhash2022optimal}
A.~Subhash, A.~Kammoun, A.~Elzanaty, S.~Kalyani, Y.~H. Al-Badarneh, and M.-S.
  Alouini, ``Optimal phase shift design for fair allocation in {RIS} aided
  uplink network using statistical {CSI},'' \emph{IEEE J.Sel. Areas Commun.},
  2023.

\bibitem{wu2019beamforming}
Q.~Wu and R.~Zhang, ``Beamforming optimization for wireless network aided by
  intelligent reflecting surface with discrete phase shifts,'' \emph{IEEE
  Trans. Commun.}, vol.~68, no.~3, pp. 1838--1851, 2019.

\bibitem{xu2020resource}
D.~Xu, X.~Yu, Y.~Sun, D.~W.~K. Ng, and R.~Schober, ``Resource allocation for
  irs-assisted full-duplex cognitive radio systems,'' \emph{IEEE Trans.
  Commun.}, vol.~68, no.~12, pp. 7376--7394, 2020.

\bibitem{masmoudi2015maximum}
A.~Masmoudi and T.~Le-Ngoc, ``A maximum-likelihood channel estimator for
  self-interference cancelation in full-duplex systems,'' \emph{IEEE Trans.
  Veh. Technol.}, vol.~65, no.~7, pp. 5122--5132, 2015.

\bibitem{muranov2021deep}
K.~Muranov, M.~A. Islam, B.~Smida, and N.~Devroye, ``On deep learning assisted
  self-interference estimation in a full-duplex relay link,'' \emph{IEEE
  Wireless Commun. Lett.}, vol.~10, no.~12, pp. 2762--2766, 2021.

\bibitem{raj2022deep}
V.~Raj, N.~Nayak, and S.~Kalyani, ``Deep reinforcement learning based blind
  mmwave {MIMO} beam alignment,'' \emph{IEEE Trans. Wireless Commun.}, vol.~21,
  no.~10, pp. 8772--8785, 2022.

\bibitem{huang2020reconfigurable}
C.~Huang, R.~Mo, and C.~Yuen, ``Reconfigurable intelligent surface assisted
  multiuser {MISO} systems exploiting deep reinforcement learning,'' \emph{IEEE
  J.Sel. Areas Commun.}, vol.~38, no.~8, pp. 1839--1850, 2020.

\bibitem{huang2021multi}
C.~Huang, Z.~Yang, G.~C. Alexandropoulos, K.~Xiong, L.~Wei, C.~Yuen, Z.~Zhang,
  and M.~Debbah, ``Multi-hop ris-empowered terahertz communications: A
  drl-based hybrid beamforming design,'' \emph{IEEE Journal on Selected Areas
  in Communications}, vol.~39, no.~6, pp. 1663--1677, 2021.

\bibitem{silver2014deterministic}
D.~Silver, G.~Lever, N.~Heess, T.~Degris, D.~Wierstra, and M.~Riedmiller,
  ``Deterministic policy gradient algorithms,'' in \emph{International
  conference on machine learning}, ser. ICML'14.\hskip 1em plus 0.5em minus
  0.4em\relax JMLR.org, 2014, p. I–387–I–395.

\bibitem{lillicrap2015continuous}
T.~P. Lillicrap, J.~J. Hunt, A.~Pritzel, N.~Heess, T.~Erez, Y.~Tassa,
  D.~Silver, and D.~Wierstra, ``Continuous control with deep reinforcement
  learning,'' \emph{arXiv preprint arXiv:1509.02971}, 2015.

\bibitem{glorot2010understanding}
X.~Glorot and Y.~Bengio, ``Understanding the difficulty of training deep
  feedforward neural networks,'' in \emph{Proc. 13th international conference
  on artificial intelligence and statistics}.\hskip 1em plus 0.5em minus
  0.4em\relax JMLR Workshop and Conference Proceedings, 2010, pp. 249--256.

\bibitem{mnih2013playing}
V.~Mnih, K.~Kavukcuoglu, D.~Silver, A.~Graves, I.~Antonoglou, D.~Wierstra, and
  M.~Riedmiller, ``Playing atari with deep reinforcement learning,''
  \emph{arXiv preprint arXiv:1312.5602}, 2013.

\bibitem{pan2020multicell}
C.~Pan, H.~Ren, K.~Wang, W.~Xu, M.~Elkashlan, A.~Nallanathan, and L.~Hanzo,
  ``Multicell {MIMO} communications relying on intelligent reflecting
  surfaces,'' \emph{IEEE Trans. Wireless Commun.}, vol.~19, no.~8, pp.
  5218--5233, 2020.

\bibitem{evmorfos2022deep}
S.~Evmorfos, A.~P. Petropulu, and H.~V. Poor, ``Deep reinforcement learning for
  {IRS} phase shift design in spatiotemporally correlated environments,''
  \emph{arXiv preprint arXiv:2211.09726}, 2022.

\bibitem{balanis2015antenna}
C.~A. Balanis, \emph{Antenna Theory: Analysis and Design}.\hskip 1em plus 0.5em
  minus 0.4em\relax John Wiley \& Sons, 2015.

\bibitem{jakes1994microwave}
W.~C. Jakes and D.~C. Cox, \emph{Microwave mobile communications}.\hskip 1em
  plus 0.5em minus 0.4em\relax Wiley-IEEE press, 1994.

\bibitem{5gmodel2016}
N.~Docomo \emph{et~al.}, ``5{G} channel model for bands up to100 {GH}z,'' Tech.
  Report, Oct, Tech. Rep., 2016.

\bibitem{rappaport2010wireless}
T.~S. Rappaport, \emph{Wireless communications: Principles and practice,
  2/E}.\hskip 1em plus 0.5em minus 0.4em\relax Pearson Education India, 2010.

\bibitem{riihonen2011hybrid}
T.~Riihonen, S.~Werner, and R.~Wichman, ``Hybrid full-duplex/half-duplex
  relaying with transmit power adaptation,'' \emph{IEEE Trans. Wireless
  Commun.}, vol.~10, no.~9, pp. 3074--3085, 2011.

\bibitem{suraweera2014low}
H.~A. Suraweera, I.~Krikidis, G.~Zheng, C.~Yuen, and P.~J. Smith,
  ``Low-complexity end-to-end performance optimization in {MIMO} full-duplex
  relay systems,'' \emph{IEEE Trans. Wireless Commun.}, vol.~13, no.~2, pp.
  913--927, 2014.

\bibitem{wen2018deep}
C.-K. Wen, W.-T. Shih, and S.~Jin, ``Deep learning for massive {MIMO} {CSI}
  feedback,'' \emph{IEEE Wireless Commun. Lett.}, vol.~7, no.~5, pp. 748--751,
  2018.

\bibitem{sun2014mimo}
S.~Sun, T.~S. Rappaport, R.~W. Heath, A.~Nix, and S.~Rangan, ``{MIMO} for
  millimeter-wave wireless communications: Beamforming, spatial multiplexing,
  or both?'' \emph{IEEE Communications Magazine}, vol.~52, no.~12, pp.
  110--121, 2014.

\bibitem{jeon2017new}
Y.~Jeon, C.~Song, S.-R. Lee, S.~Maeng, J.~Jung, and I.~Lee, ``New beamforming
  designs for joint spatial division and multiplexing in large-scale miso
  multi-user systems,'' \emph{IEEE Trans. Wireless Commun.}, vol.~16, no.~5,
  pp. 3029--3041, 2017.

\end{thebibliography}

\end{document}